\documentclass[10pt,journal,compsoc]{IEEEtran}

\ifCLASSOPTIONcompsoc
  \usepackage{cite}
\else
  \usepackage{cite}
\fi

\usepackage{cite}

\ifCLASSINFOpdf

\else
\fi
\usepackage{courier}
\usepackage{graphicx}
\usepackage[T1]{fontenc} 
\usepackage{amsmath}
\usepackage{mathrsfs}
\usepackage[cmintegrals]{newtxmath}
\usepackage{bm} 
\usepackage{amsmath}
\usepackage{graphicx}
\usepackage{algorithm}
\usepackage{algorithmicx}
\usepackage{algpseudocode}
\usepackage{array}
\usepackage{tabularx}
\usepackage{balance}
\usepackage{makecell}
\usepackage{color}
\usepackage{multirow}
\usepackage{tabularx}
\usepackage{amsfonts}
\usepackage{mathrsfs}
\usepackage{cite}

\usepackage{xpatch}



\ifCLASSOPTIONcompsoc
\usepackage[caption=false,font=normalsize,labelfon
t=sf,textfont=sf]{subfig}
\else
\usepackage[caption=false,font=footnotesize]{subfi
g}
\fi

\makeatletter
\newcommand{\multiline}[1]{%
  \begin{tabularx}{\dimexpr\linewidth-\ALG@thistlm}[t]{@{}X@{}}
    #1
  \end{tabularx}
}
\makeatother

\DeclareMathOperator*{\argmax}{arg\,max}

\begin{document}
%
\title{Fast Adaptive Task Offloading in Edge Computing based on Meta Reinforcement Learning}
%
%
%
%

\author{Jin~Wang,
        Jia~Hu,
        Geyong~Min,
        ~Albert~Y.~Zomaya, \textit{Fellow, IEEE}, 
        and~Nektarios~Georgalas
\IEEEcompsocitemizethanks{\IEEEcompsocthanksitem Jin Wang, Jia Hu, and Geyong Min are with the Department
of Computer Science, University of Exeter, United Kingdom.\protect\\
E-mail: \{jw855, j.hu, g.min\}@exeter.ac.uk 
\IEEEcompsocthanksitem Albert Y. Zomaya is with the School of Information Technologies, The University of Sydney, Australia. \protect \\
E-mail: albert.zomaya@sydney.edu.au
\IEEEcompsocthanksitem Nektarios Georgalas is with Applied Research Department, British Telecom, United Kingdom. \protect \\
E-mail: nektarios.georgalas@bt.com\protect 
\IEEEcompsocthanksitem Corresponding authors: Jia Hu and Geyong Min. \protect
\IEEEcompsocthanksitem Accepted by IEEE Transactions on Parallel and Distributed Systems
\IEEEcompsocthanksitem The source code of MRLCO implementation can be found online: https://github.com/linkpark/metarl-offloading
}
\thanks{}}

%

\makeatletter
\long\def\@IEEEtitleabstractindextextbox#1{\parbox{0.922\textwidth}{#1}} 
\makeatother

\IEEEtitleabstractindextext{%
\begin{abstract}

Multi-access edge computing (MEC) aims to extend cloud service to the network edge to reduce network traffic and service latency. A fundamental problem in MEC is how to efficiently offload heterogeneous tasks of mobile applications from user equipment (UE) to MEC hosts. Recently, many deep reinforcement learning (DRL) based methods have been proposed to learn offloading policies through interacting with the MEC environment that consists of UE, wireless channels, and MEC hosts. However, these methods have weak adaptability to new environments because they have low sample efficiency and need full retraining to learn updated policies for new environments. To overcome this weakness, we propose a task offloading method based on meta reinforcement learning, which can adapt fast to new environments with a small number of gradient updates and samples. We model mobile applications as Directed Acyclic Graphs (DAGs) and the offloading policy by a custom sequence-to-sequence (seq2seq) neural network. To efficiently train the seq2seq network, we propose a method that synergizes the first order approximation and clipped surrogate objective. The experimental results demonstrate that this new offloading method can reduce the latency by up to 25\% compared to three baselines while being able to adapt fast to new environments.
\end{abstract}

}

\maketitle
\IEEEdisplaynontitleabstractindextext

%
\IEEEpeerreviewmaketitle

\IEEEraisesectionheading{\section{Introduction}\label{sec:introduction}}

\IEEEPARstart{R}ecent years have witnessed the rapid advance of new computing and communication technologies, driving the increasing emergence of innovative mobile applications and services, such as augmented reality, virtual reality, face recognition, and mobile healthcare. These mobile applications introduce a significant surge in demands for computing and storage resources that are often provided by cloud servers. This situation generates huge network traffic between cloud and users, thus placing a heavy burden on the backhaul links and causing high service latency. Multi-access Edge Computing (MEC) \cite{reznik2019developing} was recently introduced as a key technology to address this problem. The underlying principle of MEC is to extend cloud computing capabilities to MEC host at the network edge close to users, which can significantly alleviate network congestion and reduce service latency. 


One of the key functionalities of MEC is task offloading (aka, computation offloading), which enables to offload computation-intensive tasks of mobile applications from user equipment (UE) to MEC host at the network edge. In real-world scenarios, many mobile applications (e.g., face recognition \cite{ra2011odessa}, gesture recognition \cite{ra2011odessa}, and augmented reality\cite{al2017energy}) are composed of dependent tasks, which can be modelled as a Directed Acyclic Graph (DAG). Thus, offloading dependent tasks in a DAG with the minimum latency is a crucial problem in MEC. Since this problem is NP-hard, many existing solutions are based on heuristic or approximation algorithms \cite{lin2015task,dinh2017offloading,wu2019efficient}. However, these solutions rely heavily on expert knowledge or accurate mathematical models for the MEC system. Whenever the environment of the MEC system changes, the expert knowledge or mathematical models may need to be updated accordingly. Therefore, it is difficult for one specific heuristic/approximation algorithm to fully adapt to the dynamic MEC scenarios arisen from the increasing complexity of applications and architectures of MEC.

Deep Reinforcement Learning (DRL), which combines reinforcement learning (RL) with Deep Neural Network (DNN), provides a promising solution to the above challenge, because DRL can learn to solve complex problems such as games \cite{silver2017mastering}, robotics \cite{lillicrap2015continuous}, and traffic scheduling \cite{chinchali2018cellular} by trial and error without accurate models for the environment. More recently, researchers studied the application of DRL to various MEC task offloading problems \cite{dinh2018learning,chen2019optimized,wang2019computation,huang2019deeprl}. They considered the MEC system including UE, wireless channels, and MEC host as one stationary RL environment and learn an offloading policy through interacting with the environment. However, these methods have weak adaptability for unexpected perturbations or unseen situations (i.e., new environments) like changes of applications, task numbers, or data rates. Because they have low sample efficiency and need full retraining to learn an updated policy for the new environment, they are time-consuming. 


Meta learning \cite{vanschoren2018meta} is a promising method to address the aforementioned issues by leveraging previous experiences across a range of learning tasks to significantly accelerate learning of new tasks. In the context of RL problems, meta reinforcement learning (MRL) aims to learn policies for new tasks within a small number of interactions with the environment by building on previous experiences. In general, MRL conducts two ``loops'' of learning, an ``outer loop'' which uses its experiences over many task contexts to gradually adjust parameters of the meta policy that governs the operation of an ``inner loop''. Based on the meta policy, the ``inner loop'' can adapt fast to new tasks through a small number of gradient updates \cite{botvinick2019reinforcement}. 

There are significant benefits of adapting MRL to solving the computation offloading problem. Firstly, specific policies for new mobile users can be fast learned based on their local data and the meta policy. Secondly, MRL training in the MEC system can leverage resources from both the MEC host and UE. More specifically, training for the meta policy (outer loop) is run on the MEC host and training for the specific offloading policy (inner loop) is processed on UE. Normally, the ``inner loop'' training only needs several training steps and a small amount of sampling data, thus the UE with limited computation resources and data is able to complete the training process. Finally, MRL can significantly improve the training efficiency in learning new tasks and make the offloading algorithm more adaptive to the dynamic MEC environment. 

In this paper, we propose an MRL-based method that synergizes the first-order MRL algorithm with a sequence-to-sequence (seq2seq) neural network. The proposed method learns a meta offloading policy for all UE and fast obtains the effective policy for each UE based on the meta policy and local data. To evaluate the performance of the MRLCO under dynamic scenarios, we consider the following scenarios: 1) Heterogeneous users with personal preferences of mobile applications which are represented as DAGs with different heights, widths, and task numbers. 2) Varying transmission rates according to the distance between the UE and the MEC host. 


The major contributions of this paper can be summarized as follows:
\begin{itemize}
   \item This paper is the first of its kind to propose an MRL-based method (MRLCO) to address the computation offloading problem, achieving fast adaptation to dynamic offloading scenarios. MRLCO has high sample efficiency towards new learning tasks, thus it enables UE to run the training process by using its own data even with limited computation resources. 

    \item We propose a new idea to model the dynamic computation offloading process as multiple MDPs, where the learning of offloading policies is decomposed into two parts: effectively learning a meta policy among different MDPs, and fast learning a specific policy for each MDP based on the meta policy.

    \item  We convert the offloading decision process as a sequence prediction process and design a custom seq2seq neural network to represent the offloading policy. An embedding method is also proposed to embed the vertices of a DAG considering both its task profiles and dependencies. In addition, we propose a new training method which combines the first-order approximation and clipped surrogate objective to stabilize the training of the seq2seq neural network.

    \item We conduct simulation experiments using generated synthetic DAGs according to real-world applications, covering a wide range of topologies, task numbers, and transmission rates. The results show that MRLCO achieves the lowest latency within a small number of training steps compared to three baseline algorithms including a fine-tuning DRL method, a greedy algorithm, and a heterogeneous earliest finish time (HEFT) based heuristic algorithm. 
\end{itemize}

The rest of the paper is organised as follows. A brief introduction to MEC, RL, and MRL is given in Section \ref{sec::background}. The problem formulation for task offloading is presented in Section \ref{sec::problem_formulation}. The details of the MRLCO are described in Section \ref{sec::mrl-taskoffloading}. Evaluation results are presented and discussed in Section \ref{sec::experiments}. The related work is reviewed in Section \ref{sec::relatedwork}. We discuss the MRLCO and its future work in Section \ref{sec::disscussion}. Finally, Section \ref{sec::conclusion} concludes the paper. 

\section{Background}
\label{sec::background}
This section briefly introduces the background related to MEC, RL, and MRL.
\subsection{Multi-access Edge Computing}
Over recent years, MEC has been acknowledged as one of the emerging network paradigms, which can release the pressure introduced by an unprecedented increase in traffic volume and computation demands nowadays through enabling cloud services to the network edge. Typically, MEC hosts coupled with computation and storage resources are deployed in the network edge, supporting intensive computation and data processing. As such, MEC can alleviate the burden of backhaul links and cut down the service latency. MEC is beneficial to a wide variety of emerging applications that require high volume data and low latency, e.g., autonomous driving, augmented reality, and digital healthcare. 

{\color{black} In practice, many mobile applications are composed of multiple tasks with inner dependencies among them, which can be offloaded to MEC hosts for processing. Specifically, the objective of task offloading is to find the optimal policy to partition an application into two groups of computation tasks with one executed on the UE and the other offloaded to an MEC host so that the total running cost is minimal. }


\subsection{Reinforcement Learning}
\label{sec::background-rl}
RL considers learning from environment so as to maximize the accumulated reward. Formally, a learning task,  $\mathcal{T}$, is modelled as an MDP, which is defined by a tuple $\left( \mathcal{S}, \mathcal{A}, \mathcal{P}, \mathcal{P}_0, \mathcal{R}, \gamma \right)$. Here, $\mathcal{S}$ is the state space, $\mathcal{A}$ denotes the action space, $\mathcal{R}$ is a reward function, $\mathcal{P}$ is the state-transition probabilities matrix, $\mathcal{P}_0$ is the initial state distribution, and $\gamma \in [0, 1]$ is the discount factor. A policy $\pi(a|s)$, where $a \in \mathcal{A}$ and $s \in \mathcal{S}$, is a mapping from state $s$ to the probability of selecting action $a$. We define the trajectories sampled from the environment according to the policy $\pi$ as $\tau_{\pi} = (s_0, a_0, r_0, s_1, a_1, r_1, \ldots)$, where $s_0 \sim \mathcal{P}_0$, $a_t \sim \pi( \cdot | s_{t})$ and $r_t$ is a reward at time step $t$. 

The state value function of a state $s_t$ under a parameterized policy $\pi(a|s; \theta)$,  denoted as $v_{\pi}(s_t)$, is the expected return when starting in $s_t$ and following $\pi(a|s; \theta)$ thereafter. Here, $\theta$ is the vector of policy parameters and $v_{\pi}(s_t)$ can be calculated by 
\begin{equation}
\label{state_value_func}
v_{\pi}(s_t)=\mathbb{E}_{\tau \sim P_{\mathcal{T}}(\tau|\theta)}\left[ \sum_{k=t}\gamma^{k-t} r_k \right]
\end{equation},
where $P_{\mathcal{T}}(\tau|\theta)$ is the probability distribution of sampled trajectories based on $\pi(a|s; \theta)$. The goal for RL is to find an optimal parameterized policy $\pi(a|s; \theta^*)$ to maximize the expected total rewards $J = \sum_{s_0 \sim \mathcal{P}_0} v_{\pi}(s_0)$. 

\subsection{Meta Reinforcement Learning}
MRL enhances the conventional RL methods with meta learning, which aims to learn a learning algorithm that can quickly find the policy for a learning task $\mathcal{T}_i$ drawn from a distribution of tasks $\rho(\mathcal{T})$. Each learning task $\mathcal{T}_i$ corresponds to a different MDP, and these learning tasks typically share the same state and action spaces but may differ in reward functions or their dynamics (i.e., $\mathcal{P}$ and $\mathcal{P}_0$). 


Recent years have brought a wealth of methods focused on different aspects of MRL. One typical example is gradient-based MRL, which aims to learn the initial parameters $\theta$ of a policy neural network, so that performing a single or few steps of policy gradient over $\theta$ with a given new task can lead to an effective policy for that task. We follow the formulation of model-agnostic meta-learning (MAML) \cite{finn2017model}, giving the target of gradient-based MRL as 
\begin{equation}
\label{MRL-obj}
\begin{aligned}
&J(\theta) = \mathbb{E}_{\mathcal{T}_i \sim \rho(\mathcal{T})} \left [  J_{\mathcal{T}_i}(\theta')  \right ], {\rm with} \ \theta' := U(\theta, \mathcal{T}_i),
\end{aligned}
\end{equation}
where $J_{\mathcal{T}_i}$ denotes the objective function of task $\mathcal{T}_i$. For example, when using vanilla policy gradient (VPG), $J_{\mathcal{T}_i}(\theta) = \mathbb{E}_{ \tau \sim P_{\mathcal{T}_i}(\tau|\theta)} \sum_{t=0}\left( \gamma^t r_t - b(s_t)\right )$, where $b(s_t)$ denotes an arbitrary baseline which does not vary with $a_t$. $U$ denotes the update function which depends on $J_{\mathcal{T}_i}$ and the optimization method. For instance, if we conduct $k$-step gradient ascent for $\mathcal{T}_i$, then $U(\theta, \mathcal{T}_i) = \theta + \alpha\sum_{t=1}^k g_t$, where $g_t$ denotes the gradient of $J_{\mathcal{T}_i}$ at time step $t$ and $\alpha$ is the learning rate. Therefore, the optimal parameters of policy network and update rules are
\begin{equation}
\label{MRL-update}
\begin{aligned}
& \theta^{*} = \argmax_{\theta} \mathbb{E}_{\mathcal{T}_i \sim \rho(\mathcal{T})} \left [  J_{\mathcal{T}_i}(U(\theta, \mathcal{T}_i)  \right ], \\
&\theta \leftarrow \theta + \beta \mathbb{E}_{\mathcal{T}_i \sim \rho(\mathcal{T})} \left [ \nabla_{\theta}J_{\mathcal{T}_i}(U(\theta, \mathcal{T}_i)  \right ],
\end{aligned}
\end{equation}
where $\beta$ is the learning rate of ``outer loop'' training. 
The gradient-based MRL has good generalization ability. However, the second-order derivative in MAML may bring huge computation cost during training, which is inefficient. In addition, when combining with a complex neural network architecture, e.g., a seq2seq neural network, the implementation of second-order MAML becomes intractable. To address these challenges, some  algorithms \cite{finn2017model,nichol2018first} use the first-order approximation to MAML target. In this work, we implement MRLCO based on the first-order MRL due to its low computation cost, good performance, and easy implementation when combing with a seq2seq neural network.

\begin{figure}[t]
    \centering
    \includegraphics[width=3.0in]{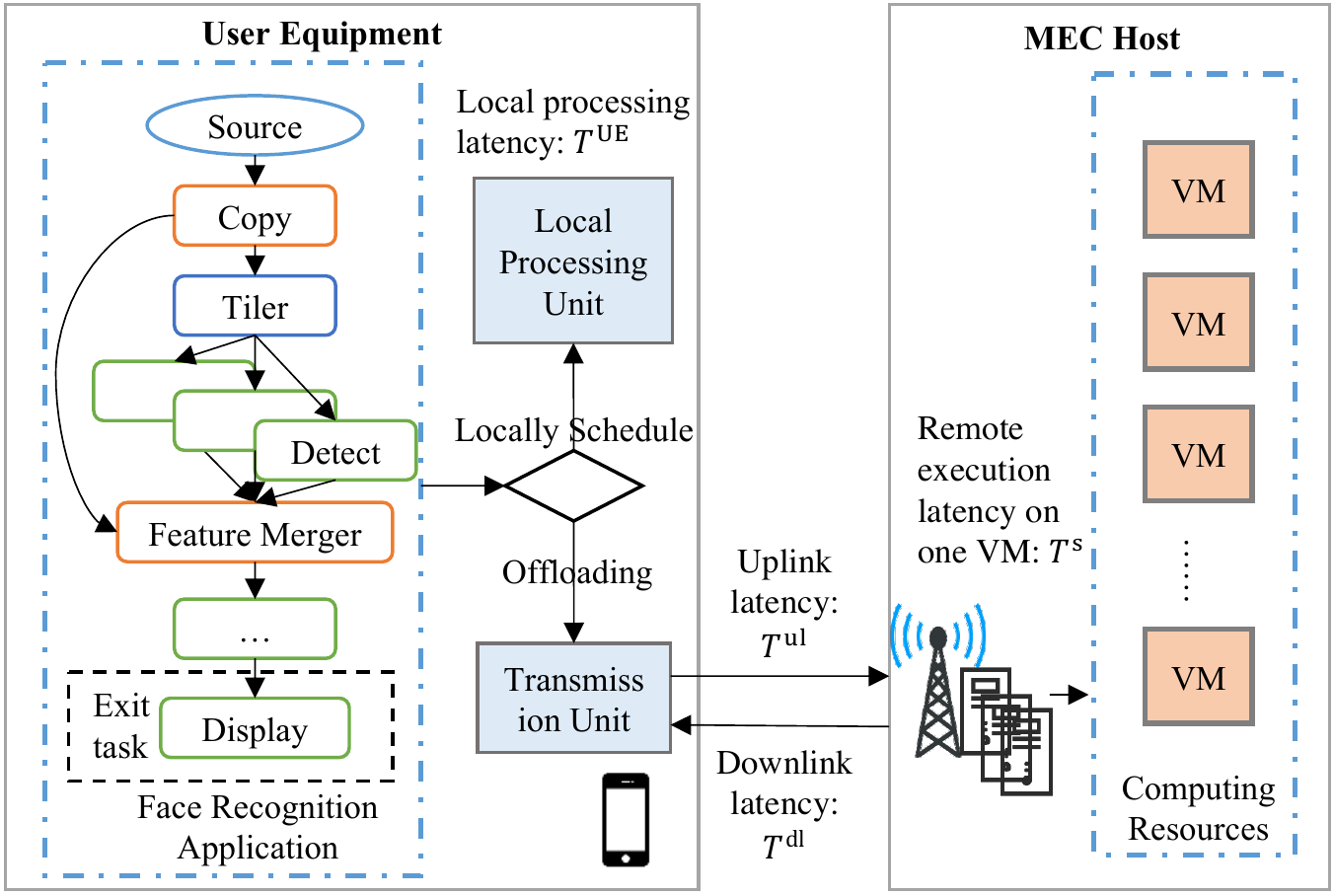}
    \centering\caption{An example of computation offloading in MEC.}
    \label{mec_example}
\end{figure}

\section{Problem Formulation}
\label{sec::problem_formulation}
Fig. \ref{mec_example} gives an example of computation offloading in MEC. This example considers a real-world application---face recognition, which consists of dependent tasks such as tiler, detection, or feature mergence \cite{ra2011odessa}. The UE makes offloading decisions for those tasks according to the system status and task profiles, thus some tasks are run locally on the UE while others are offloaded to the MEC host via wireless channels. In general, each MEC host runs multiple virtual machines (VMs) processing the tasks. In this work, we consider that each UE is associated with a dedicated VM providing private computing, communications and storage resources to the UE, the same as in works \cite{sun2017avaptive,borcea2015avatar}. {\color{black} The computation capacity (i.e., the number of CPU cores times the clock speed of each core) of an MEC host is denoted as $f_{s}$. We consider an equal resource allocation for VMs, i.e., all VMs evenly share the computing resource of the MEC host. Therefore, assuming there are $k$ users in the MEC systems, the computation capacity for each VM is $f_{\rm vm} = f_{s} / k$.} Formally, we model mobile applications as DAGs, $G=(T, E)$, where the vertex set $T$ represents the tasks and the directed edge set $E$ represents the dependencies among tasks, respectively. Each directed edge is denoted by $\overrightarrow{e} = (t_i, t_j)$, corresponding to the dependency between task $t_i$ and $t_j$, where $t_i$ is an immediate parent task of $t_j$, and $t_j$ is an immediate child task of $t_i$. With the constraint of dependency, a child task cannot be executed until all of its parent tasks are completed. In $G=(T, E)$, we call a task without any child task as an \textit{exit task}.   

In computation offloading, a computation task can either be offloaded to the MEC host or executed locally on the UE. If task $t_i$ is offloaded, there are three steps to execute $t_i$. First, the UE sends $t_i$ to an MEC host through a wireless channel. Second, the MEC host runs the received task. Finally, the running result of $t_i$ is returned to the UE. The latency at each step is related to the task profile and the MEC system state. Here, the task profile of $t_i$ includes required CPU cycles for running the task, $C_i$, data sizes of the task sent, ${\rm data}_{i}^{s}$, and the result received, ${\rm data}_{i}^{r}$. Besides, the MEC system state contains the transmission rate of wireless uplink channel, $R_{\rm ul}$, and rate of downlink channel, $R_{\rm dl}$. Therefore, the latency for sending data, $T_{i}^{\rm ul}$, executing on the MEC host, $T_{i}^{\rm s}$, and receiving result, $T_{i}^{\rm dl}$, of task $t_i$ can be calculated as:
\begin{equation}
    \label{offloading_latency}
    {T_{i}^{\rm ul} = {\rm data}_{i}^{s} / R_{\rm ul}, \ \  T_{i}^{\rm s} = C_i / f_{\rm vm}, \ \  T_{i}^{\rm dl} = {\rm data}_{i}^{r} / R_{\rm dl}}.
\end{equation}
If task $t_i$ runs locally on the UE, there is only running latency on the UE, which can be obtained by $T_{i}^{\rm UE} = C_i / f_{\rm UE}$ where $f_{\rm UE}$ denotes the computation capacity of the UE. {\color{black} The end-to-end latency of a task offloading process includes local processing, uplink, downlink, and remote processing latency, as shown in Fig. \ref{mec_example}.}

The scheduling plan for a DAG, $G=(T, E)$, is denoted as $A_{1:n} = \{ a_1, a_2, \ldots, a_n \}$, where $|T| = n$ and $a_i$ represents the offloading decision of $t_i$. Tasks are scheduled in a sequence based on the scheduling plan, where all parent tasks are scheduled before their child tasks. We denote $FT^{\rm ul}_{i}$, $FT^{\rm s}_{i}$, $FT^{\rm dl}_{i}$, and $FT^{\rm UE}_{i}$ as the finish time of task $t_i$ on the uplink wireless channel, the MEC host, the downlink wireless channel, and the UE, respectively. We also denote the available time of these resources when scheduling task $t_i$ as $\mathcal{M}^{\rm ul}_{i}$, $\mathcal{M}^{\rm s}_{i}$, $\mathcal{M}^{\rm dl}_{i}$, and $\mathcal{M}^{\rm UE}_{i}$. The resource available time depends on the finish time of the task scheduled immediately before $t_i$ on that resource. If the task scheduled immediately before $t_i$ does not utilize the resource, we set the finish time on the resource as 0. 

If task $t_i$ is offloaded to the MEC host, $t_i$ can only start to send its data when its parent tasks are all completed and the uplink channel is available. Therefore, the finish time on the uplink channel, $FT^{\rm ul}_{i}$, can be defined by 
\begin{equation}
\label{uplink_channel_ft}
\begin{aligned}
&FT^{\rm ul}_{i} = \max \left \{ \mathcal{M}^{\rm ul}_{i}, \max_{j \in {\rm parent}(t_i)} \left \{ FT^{\rm UE}_{j}, FT^{\rm dl}_{j}  \right \} \right \} + T^{\rm ul}_{i}, \\ 
&\mathcal{M}^{\rm ul}_{i} = \max \left \{ \mathcal{M}^{\rm ul}_{i-1}, FT^{\rm ul}_{i-1} \right \}. 
\end{aligned}
\end{equation}
Similarly, the finish time of $t_i$ on the MEC host, $FT^{\rm s}_{i}$, and that on the downlink channel, $FT^{\rm dl}_{i}$, are given by
\begin{equation}
\label{server_downlink_ft}
\begin{aligned}
&FT^{\rm s}_{i} = \max \left \{ \mathcal{M}^{\rm s}_{i}, \max \left \{ FT^{\rm ul}_{i}, \max_{j \in {\rm parent}(t_i)} FT^s_j \right \} \right \} + T^{\rm s}_{i}, \\ 
&FT^{\rm dl}_{i} = \max \left \{ \mathcal{M}^{\rm dl}_{i}, FT^{\rm s}_{i} \right \} + T^{\rm dl}_{i}, \\
&\mathcal{M}^{\rm s}_{i} = \max \left \{ \mathcal{M}^{\rm s}_{i-1}, FT^{\rm s}_{i-1} \right \}, \\
&\mathcal{M}^{\rm dl}_{i} = \max \left \{ \mathcal{M}^{\rm dl}_{i-1}, FT^{\rm dl}_{i-1} \right \}. 
\end{aligned}
\end{equation}

If $t_i$ is scheduled on the UE, the start time of $t_i$ depends on the finish time of its parent tasks and the available time of the UE. Formally, the finish time of $t_i$ on the UE, $FT^{\rm UE}_{i}$, is defined as
\begin{equation}
\label{ue_ft}
\begin{aligned}
FT^{\rm UE}_{i} = &\max \left \{ \mathcal{M}^{\rm UE}_i, \max_{j \in {\rm parent}(t_i)} \left \{ FT^{\rm UE}_{j}, FT^{\rm dl}_{j}  \right \} \right \} + T^{\rm UE}_{i}, \\ 
\mathcal{M}^{\rm UE}_{i} = &\max \left \{ \mathcal{M}^{\rm UE}_{i-1}, FT^{\rm UE}_{i-1} \right \}. 
\end{aligned}
\end{equation}

Overall, the objective is to find an effective offloading plan for the DAG to obtain the minimal total latency. Formally, the total latency of a DAG given a scheduling plan $A_{1:n}$, $T^{c}_{A_{1:n}}$, is given by
\begin{equation}
\label{object_function}
    T^{c}_{A_{1:n}} = \max \left[ \max_{t_k \in \mathcal{K}} \left( FT^{\rm UE}_k, FT^{\rm dl}_k \right ) \right ],
\end{equation}
where $\mathcal{K}$ is the set of exit tasks. The problem in Eq. (\ref{object_function}) is NP-hard, so finding the optimal offloading plan can be extremely challenging due to the highly dynamic DAG topologies and MEC system states. In the next section, we present the details of MRLCO for handling this problem. 


\begin{table}[t]
\renewcommand\arraystretch{1.2}
 \renewcommand{\tabcolsep}{0.7mm}
 \caption{Summary of Main Notations}
 \label{tb:simulation_para}
 \centering
 \begin{tabular}{ >{\centering\arraybackslash}p{0.8in} | p{2.5in}  }
  \hline
  \textbf{Notation} & \textbf{Description} \\ 
  \hline
  $\mathbb{E}$ & Mean value  \\
  \hline
  ${\rm data}_i^s$, ${\rm data}_i^r$ & Size of data sending to or receiving from a task $t_i$\\
  \hline
  $R_{\rm ul}, R_{\rm dl}$ & Transmission rate of uplink and downlink  \\
  \hline
  $f_{\rm UE}$, $f_{\rm s}$, $f_{\rm vm}$  & Computation capacity of  UE, MEC host and VM \\
  \hline
  $T_{i}^{\rm ul}$, $T_{i}^{\rm s}$, $T_{i}^{\rm dl}$, $T_{i}^{\rm UE}$ & Latency for task $t_i$ on uplink channel, MEC host, downlink channel, and UE. \\
  \hline
  $FT_{i}^{\rm ul}$, $FT_{i}^{\rm s}$, $FT_{i}^{\rm dl}$, $FT_{i}^{\rm UE}$ & Finish time for task $t_i$ on uplink channel, MEC host, downlink channel, and UE \\
  \hline
  $\mathcal{M}_{i}^{\rm ul}$, $\mathcal{M}_{i}^{\rm s}$, $\mathcal{M}_{i}^{\rm dl}$, $\mathcal{M}_{i}^{\rm UE}$ & Resource available time for task $t_i$ on uplink channel, MEC host, downlink channel, and UE  \\
  \hline
  $A_{1:n}$  & Computational offloading plan for $n$ tasks \\
  \hline
  $\mathcal{T}_i, \rho(\mathcal{T})$  & A learning task and distribution of learning tasks \\
  \hline
  $s_i$, $a_i$, $r_i$ & State, action, and reward of an MDP at time step $i$ \\
  \hline
  $\pi(a|s;\theta)$, $v(s;\theta)$ & Parametrized policy and value function for computation offloading. \\
  \hline
  $\tau_\pi$ & Trajectories sampled from the environment based on the policy $\pi$. \\
  \hline
  $f_{enc}, f_{dec}$ & Functions of encoder and decoder \\
  \hline
  $e_i, d_i$ & Output of encoder and decoder at time step $i$\\
  \hline
  $c_i$ & Context vector at decoding step $i$ \\
  \hline
  $\hat{A}_t$ & Advantage function at time step t  \\
  \hline
  $U(\theta, \mathcal{T}_i)$ & Update function (e.g., Adam) for the learning task \\
  \hline
 \end{tabular}
\end{table}

\section{MRLCO: An MRL-based Computation Offloading Solution}
\label{sec::mrl-taskoffloading}

In this section, we first give an overview of the architecture of the MRLCO and explain how it works with the MEC system. Next, we present the detailed MDP modelling for the computation offloading problem. Finally, we describe the implementation of the MRLCO algorithm. 

\subsection{The MRLCO Empowered MEC System Architecture}
\label{MRLCO_architecture}
\begin{figure*}[t]
    \centering
    \includegraphics[width=5.5in]{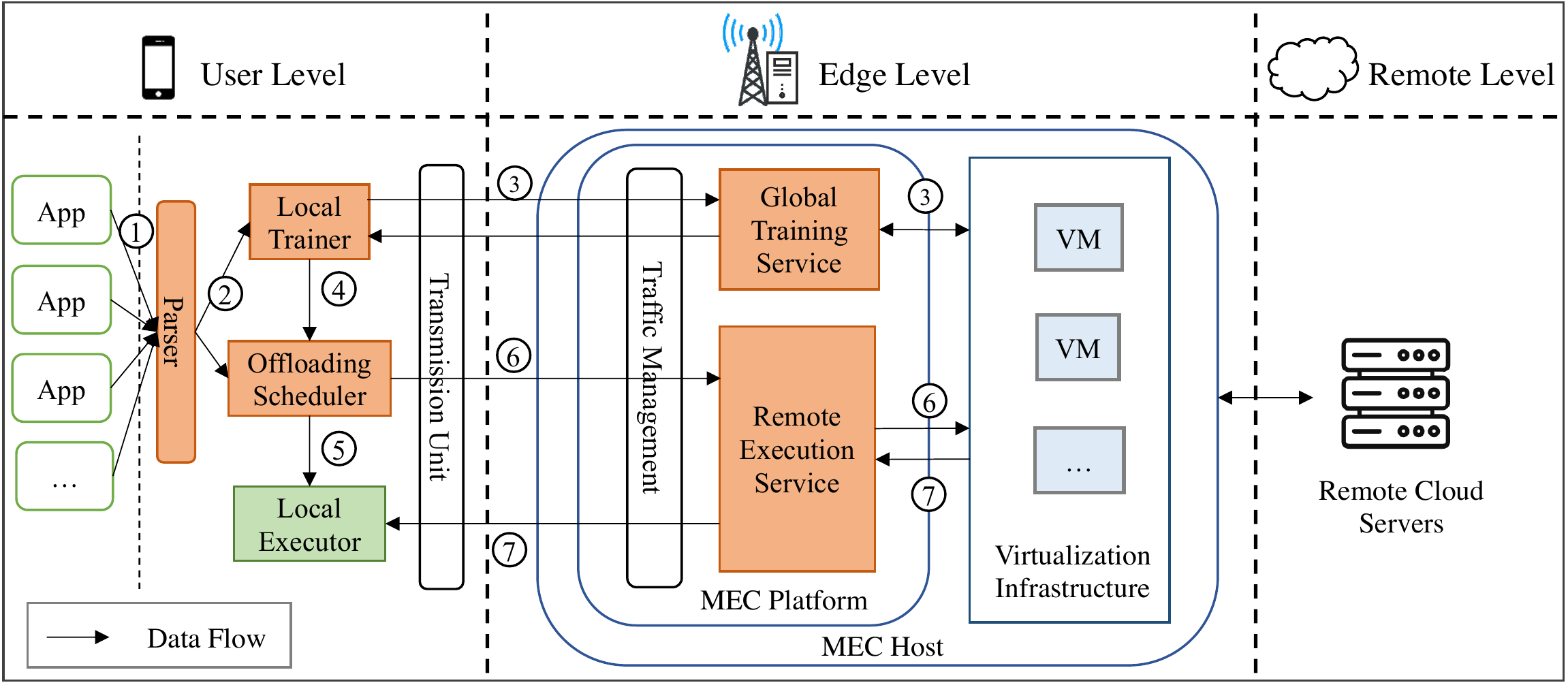}
    \centering\caption{{\color{black}The system architecture of the MRLCO empowered MEC system. The data flows in this architecture include: \textcircled{\tiny{1}} mobile applications, \textcircled{\tiny{2}} parsed DAGs, \textcircled{\tiny{3}} parameters of the policy network, \textcircled{\tiny{4}} the trained policy network, \textcircled{\tiny{5}} tasks scheduled to local executor, \textcircled{\tiny{6}} tasks offloaded to the MEC host, \textcircled{\tiny{7}} results from the offloaded tasks.}}
    \label{sys_archi}
\end{figure*}

\begin{figure*}[t]
    \centering
    \includegraphics[width=5.5in]{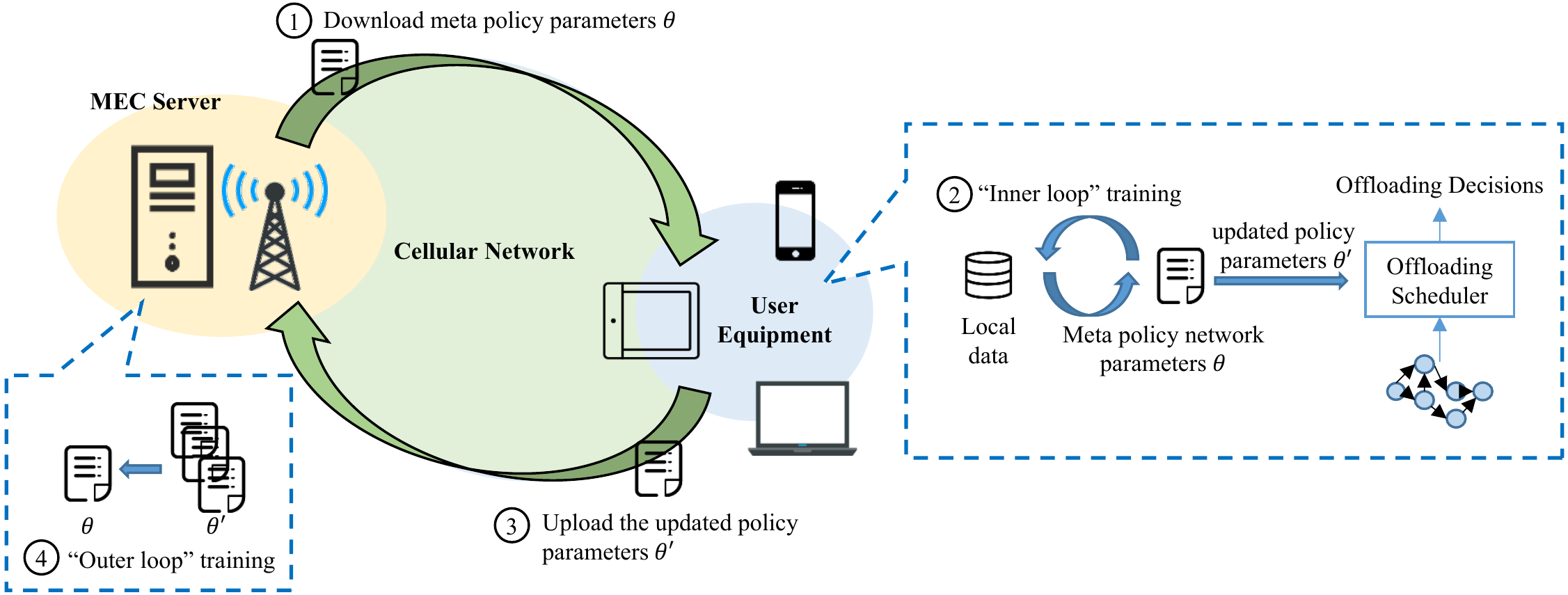}
    \centering\caption{The training process of the MRLCO empowered MEC system includes four steps: 1) the UE downloads the parameters of meta policy, $\theta$, from the MEC host; 2) ``inner loop'' training is conducted on the UE based on $\theta$ and the local data, obtaining the parameters of task-specific policy, $\theta'$; 3) the UE uploads $\theta'$ to the MEC host; 4) the MEC host conducts ``outer loop'' training based on the gathered updated parameters $\theta'$.}
    \label{mrlco_archi}
\end{figure*}

{\color{black} The MRLCO aims to leverage the computation resources from both the UE and the MEC host for the training process. There are two loops of training --- ``inner loop'' training for the task-specific policy and ``outer loop'' training for the meta policy. The ``inner loop'' training is conducted on the UE while the ``outer loop'' training on the MEC host. 

Fig. \ref{sys_archi} shows an architecture that integrates the MRLCO into an emerging MEC system \cite{reznik2019developing} composed of the user level, edge level, and remote level. Here, the user level includes heterogeneous UE, the edge level contains MEC hosts that provide edge computing services, and the remote level consists of cloud servers. Specifically, mobile users communicate with an MEC host through the local \textit{Transmission unit}. The MEC host incorporates an MEC platform and a virtualization infrastructure that provides the computing, storage, and network resources. The MEC platform provides \textit{Traffic management} (i.e., traffic rules control and domain name handling) and offers edge services. The five key modules of MRLCO (\textit{parser}, \textit{local trainer}, \textit{offloading scheduler}, \textit{global training service}, and \textit{remote execution service}) can be deployed at the user and edge levels of the MEC system separately, as described below:  
\begin{itemize}
\item At the user level, the \textit{parser} aims to convert mobile applications into DAGs. The \textit{local trainer} is responsible for the ``inner loop'' training, which receives the parsed DAGs from the \textit{parser} as training data and uploads/downloads parameters of the policy network to/from the MEC host through local transmission unit. Once the training process is finished, the trained policy network will be deployed to the \textit{offloading scheduler} that is used to make offloading decisions through policy network inference. After making decisions for all tasks of a DAG, the locally scheduled tasks will run on the local executor and the offloaded tasks will be sent to the MEC host for execution. 

\item At the edge level, the \textit{global training service} and \textit{remote execution service} modules are deployed to the MEC platform. The \textit{global training service} is used to manage the ``outer loop'' training, which sends/receives parameters of the policy network to/from the UE and deploys the global training process on the virtualization infrastructure in the MEC host. The \textit{remote execution service} is responsible for managing the tasks offloaded from the UE, assigning these tasks to associated VMs, and sending the results back to the UE. 
\end{itemize}

Next, we describe the detailed training process of the MRLCO in the MEC system, as shown in Fig. \ref{mrlco_archi}. The training process for MRLCO includes four steps. First, the UE downloads the parameters of the meta policy from the MEC host. Next, an ``inner loop'' training is run on every UE based on the meta policy and the local data, in order to obtain the task-specific policy. The UE then uploads the parameters of the task-specific policy to the MEC host. Finally, the MEC host conducts an ``outer loop'' training based on the gathered parameters of task-specific policies, generates the new meta policy, and starts a new round of training. Once obtaining the stable meta policy, we can leverage it to fast learn a task-specific policy for new UE through ``inner loop'' training. Notice that the ``inner loop'' training only needs few training steps and a small amount of data, thus can be sufficiently supported by the UE. We will present the algorithmic details of the ``outer loop'' and ``inner loop'' training in Section \ref{implement_mrlco}. }

\subsection{Modelling the Computation Offloading Process as Multiple MDPs}

To adapt MRL to solve the computation offloading problem, we firstly model the process of computation offloading under various MEC environments as multiple MDPs, where learning an effective offloading policy for one MDP is considered as a learning task. Formally, we consider a distribution over all learning tasks in MEC as $\rho(\mathcal{T})$, where each task $\mathcal{T}_i \sim \rho(\mathcal{T})$ is formulated as a different MDP, $\mathcal{T}_i=\left( \mathcal{S}, \mathcal{A}, \mathcal{P}, \mathcal{P}_0, \mathcal{R}, \gamma \right)$. (Please refer to Section \ref{sec::background-rl} for the meaning of these notations.) In order to obtain the adaptive offloading policy for all learning tasks, we decompose the learning process into two parts: effectively learning a meta policy among all MDPs and fast learning a specific offloading policy for one MDP based on the meta policy. The definitions of the state, action, and reward for the MDP are listed as follows:


\begin{itemize}
    \item \textbf{State:} When scheduling a task $t_i$, the latency of running the task depends on the task profile (i.e., required CPU cycles, data sizes), DAG topologies, the wireless transmission rate, and the state of MEC resources. According to Eqs. (\ref{uplink_channel_ft}), (\ref{server_downlink_ft}), and (\ref{ue_ft}), the state of MEC resources is related to the offloading decisions of task scheduled before $t_i$. Therefore, we define the state as a combination of the encoded DAG and the partial offloading plan:
    \begin{equation}
    \label{state_space}
    \mathcal{S} := \left \{s_i | s_i = \left ( G(T, E), A_{1:i} \right ) \right \} \ \ \ {\rm where} \ \  i \in \left [ 1, |T| \right ], 
    \end{equation} 
    where $G(T,E)$ is comprised of a sequence of task embeddings and $A_{1:i}$ is the partial offloading plan for the first $i$ tasks. To convert a DAG into a sequence of task embeddings, we first sort and index tasks according to the descending order of the rank value of each task, which is defined as
   \begin{equation}
       rank(t_i) = 
        \begin{cases}
               T_{i}^o &  \text{if $t_i \in \mathcal{K}$}, \\
               T_{i}^o + \max\limits_{t_j \in child(t_i)} \left (rank(t_j) \right )  & \text{if $t_i \notin \mathcal{K}$},
        \end{cases}
      \label{rank_value}
    \end{equation}
      where $T_{i}^o =  T_{i}^{\rm ul} + T_{i}^{\rm s} + T_{i}^{\rm dl}$ denotes the latency for task $i$ from starting offloading to finishing execution, $child(t_i)$ represents the set of immediate child tasks of $t_i$.
      Each task is converted into an embedding that consists of three elements: 1) a vector that embeds the current task index and the normalized task profile, 2) a vector that contains the indices of the immediate parent tasks, 3) a vector that contains the indices of the immediate child tasks. The size of vectors that embed parent/child task indices is limited to $p$. We pad the vector with -1, in case the number of child/parent tasks is less than $p$. 
    \item \textbf{Action: } The scheduling for each task is a binary choice, thus the action space is defined as $\mathcal{A}:=\{0, 1\}$, where 0 stands for execution on the UE and 1 represents offloading.
    \item \textbf{Reward: }The objective is to minimize $T^{c}_{A_{1:n}}$ given by Eq. (\ref{object_function}). In order to achieve this goal, we define the reward function as the estimated negative increment of the latency after making an offloading decision for a task. Formally, when taking action for the task $t_i$, the increment is defined as $\Delta T^c_{i} = T^c_{A_{1:i}} - T^c_{A_{1:i-1}}$. 
\end{itemize} 

Based on the above MDP definition, we denote the policy when scheduling $t_i$ as $\pi(a_i | G(T,E), A_{1:i-1})$. For a DAG with $n$ tasks, let $\pi \left ( A_{1:n}|G(T,E) \right )$ denote the probability of having the offloading plan $A_{1:n}$ given the graph $G(T,E)$. Therefore, $\pi(A_{1:n} | G(T,E))$ can be obtained by applying chain rules of probability on each $\pi(a_i | G(T,E), A_{1:i-1})$ as 
\begin{equation}
\label{policy_formulation}
{\pi(A_{1:n}|G(T,E)) = \prod_{i=1}^{n} \pi(a_i|G(T,E), A_{1:i-1})}.
\end{equation}

\begin{figure}[t]
    \centering
    \includegraphics[width=3.5in]{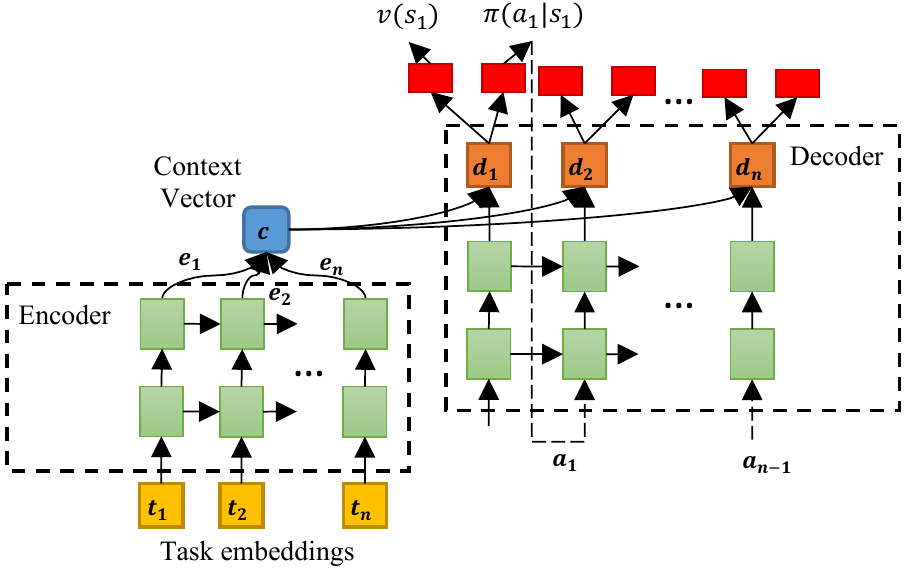}
    \centering\caption{Architecture of the seq2seq neural network in MRLCO. The architecture consists of an encoder and a decoder, where the input of the encoder is the sequence of task embeddings and the output of the decoder is used to generate both policy and value function. }
    \label{seq2seq_archi}
\end{figure}

A seq2seq neural network \cite{bahdanau2014neural} is a natural choice to represent the policy defined in Eq. (\ref{policy_formulation}). 
Fig. \ref{seq2seq_archi} shows our design of a custom seq2seq neural network, which can be divided into two parts: encoder and decoder. In our work, both encoder and decoder are implemented by recurrent neural networks (RNN). The input of the encoder is the sequence of task embeddings, $[t_1, t_2, ... , t_n]$, while the output of the decoder is the offloading decisions of each tasks, $[a_1, a_2, ..., a_n]$. To improve the performance, we include the attention mechanism \cite{bahdanau2014neural} into our custom seq2seq neural network. Attention mechanism allows the decoder to attend to different parts of the source sequence (i.e., the input sequence of the encoder) at each step of the output generation, thus it can alleviate the issue of information loss caused by the original seq2seq neural network that encodes the input sequence into a vector with fixed dimensions.

Formally, we define the functions of the encoder and decoder as $f_{enc}$ and $f_{dec}$, respectively. In our work, we use the Long Short-Term Memory (LSTM) as $f_{enc}$ and $f_{dec}$.  At each step of encoding, the output of the encoder, $e_i$, is obtained by
\begin{equation}
\label{encode_equation}
e_i = f_{enc}(t_i, e_{i-1}).
\end{equation}
After encoding all the input task embeddings, we have the output vector as $\textbf{e} = [e_1, e_2, ..., e_n]$. At each decoding step, we define the output of the decoder, $d_j$, as 
\begin{equation}
\label{decode_equation}
d_j = f_{dec}(d_{j-1}, a_{j-1}, c_j),
\end{equation}
where $c_j$ is the context vector at decoding step $j$ and is computed as a weighted sum of the outputs of the encoder:
\begin{equation}
\label{context_vector}
c_j = \sum_{i=0}^n{\alpha_{ji}e_i}.
\end{equation}
The weight $\alpha_{ji}$ of each output of encoder, $e_i$, is computed by
\begin{equation}
\alpha_{ji} = \frac{ {\rm exp} \left ( {\rm score}(d_{j-1}, e_i) \right ) }{ \sum_{k=1}^n {\rm exp} \left ( {\rm score}(d_{j-1}, e_k) \right ) },
\end{equation}
where the score function, ${\rm score}(d_{j-1}, e_i)$, is used to measure how well the input at position $i$ and the output at position $j$ match. We define the score function as a trainable feedforward neural network according to the work \cite{bahdanau2014neural}. We use the seq2seq neural network to approximate both policy $\pi(a_j| s_j)$ and value function $v_{\pi}(s_j)$ by passing the output of decoder $\textbf{d} = [d_1, d_2, ..., d_n]$ to two separate fully connected layers. Notice that the policy and value function share most of the parameters (i.e., the encoder and decoder) which are used to extract common features of DAGs (e.g., the graph structure and task profiles). Therefore, training the policy can accelerate the training of value function and vice versa. During training for the seq2seq neural network, the action $a_j$ is generated through sampling from the policy $\pi(a_j|s_j)$. Once the training is finished, the offloading decisions for a DAG can be made by inference through the seq2seq neural network, where the action $a_j$ is generated by $a_j = \argmax_{a_j} \pi(a_j|s_j)$. Therefore, the time complexity for our algorithm is the same as the inference of the seq2seq neural network with attention, which is $O(n^2)$ \cite{vinyals2015pointer}. Normally, the task number, $n$, of a mobile application is less than 100 \cite{mahmoodi2019optimal,lin2015task,dinh2017offloading}, thus the time complexity of the MRLCO is feasible. 


\subsection{Implementation of MRLCO}
\label{implement_mrlco}
MRLCO shares a similar algorithm structure with gradient-based MRL algorithms, which consists of two loops for training. Instead of using VPG as the policy gradient method for the ``inner loop'' training \cite{finn2017model}, we define our objective function based on Proximal Policy Optimization (PPO) \cite{schulman2017proximal}. Compared to VPG, PPO achieves better exploring ability and training stability. For one learning task $\mathcal{T}_i$, PPO generates trajectories using the sample policy $\pi_{\theta^o_i}$ and updates the target policy $\pi_{\theta_i}$ for several epochs, where $\theta_i$ equals $\theta^o_i$ at the initial epoch. In order to avoid a large update of the target policy, PPO uses a clipped surrogate objective as
\begin{equation}
\label{clip_target}
J_{\mathcal{T}_i}^{\rm C}(\theta_i) = \mathbb{E}_{\tau \sim P_{\mathcal{T}_i}(\tau, \theta^o_i)} \left[ \sum_{t=1}^{n} \min \left({\rm Pr}_t\hat{A}_t, {\rm clip}^{1+\epsilon}_{1-\epsilon}\left ({\rm Pr}_t  \right) \hat{A}_t \right ) \right].
\end{equation}
Here, $\theta^o_i$ is the vector of parameters of the sample policy network. ${\rm Pr}_t$ is the probability ratio between the sample policy and target policy, which is defined as
\begin{equation}
\label{clip_ratio}
{\rm Pr}_t = \frac{\pi_{\theta_i}(a_t | G(T,E), A_{1:t})}{\pi_{\theta^o_i}(a_t | G(T,E), A_{1:t})}.
\end{equation}
The clip function ${\rm clip}^{1+\epsilon}_{1-\epsilon}\left ({\rm Pr}_t \right )$ aims to limit the value of ${\rm Pr}_t$, in order to remove the incentive for moving ${\rm Pr}_t$ outside of the interval $[1-\epsilon, 1+\epsilon]$. $\hat{A}_t$ is the advantage function at time step $t$. Specially, we use general advantage estimator (GAE) \cite{schulman2015high} as our advantage function, which is defined by
\begin{equation}
\label{gae_function}
{\hat{A}_t = \sum_{k=0}^{n-t+1}(\gamma \lambda)^k (r_{t+k} + \gamma v_{\pi}(s_{t+k+1}) - v_{\pi}(s_{t+k})),}
\end{equation}
where $\lambda \in [0, 1]$ is used to control the trade-off between bias and variance. The value function loss is defined as
\begin{equation}
\label{value_loss}
J^{\rm VF}_{\mathcal{T}_i}(\theta_i) = \mathbb{E}_{\tau \sim P_{\mathcal{T}_i}(\tau, \theta^o_i)} \left[ \sum_{t=1}^{n}{\left ( v_{\pi}(s_t) - \hat{v}_{\pi}(s_t) \right )^2} \right],
\end{equation}
where $\hat{v}_{\pi}(s_t) = \sum_{k=0}^{n-t+1}\gamma^k r_{t+k}$. 

\begin{algorithm}[t]
  \caption{: Meta Reinforcement Learning based Computation Offloading}
  \label{mrlco_algo}
  \hspace*{0.02in} {\bf Require:} Task distribution $\rho(\mathcal{T})$, 
  \begin{algorithmic}[1]
    \State Randomly initialize the parameters of meta policy, $\theta$
    \For {iterations $k \in \{ 1, ..., K \}$}
    \State Sample $n$ tasks $\left \{ \mathcal{T}_0, \mathcal{T}_1, ..., \mathcal{T}_n \right \}$ from $\rho(\mathcal{T})$
    \For {each task $\mathcal{T}_i \in \left \{ \mathcal{T}_0, \mathcal{T}_1, ..., \mathcal{T}_n \right \}$}
    \State Initialize $\theta^o_i \leftarrow \theta$ and $\theta_i \leftarrow \theta$
    \State \multiline{Sample trajectories set $D = {(\tau_1, \tau_2, \ldots )}$ from $\mathcal{T}_i$ using sample policy $\pi_{\theta^o_i}$}
    \For {iterations $j \in \{ 1, ..., m \}$}
    \State \multiline{Update parameters $\theta_i$ \\ 
    $\theta_{i} \leftarrow \theta_i + \alpha \nabla_{\theta_i}J^{\rm PPO}_{\mathcal{T}_i}(\theta_i)$ \\
    by mini-batch gradient descent based on $D$ with  \textit{Adam}}
    \EndFor
    \State Update $\theta'_{i} \leftarrow \theta_{i}$
    \EndFor
    \State \multiline{Update $\theta \leftarrow \theta + \beta g^{\rm MRLCO}$ via \textit{Adam}}
    \EndFor
  \end{algorithmic}
\end{algorithm}

Overall, we combine Eq. (\ref{clip_target}) and Eq. (\ref{value_loss}), defining the objective function for each ``inner loop'' task learning as:
\begin{equation}
\label{ppo_target}
{J^{\rm PPO}_{\mathcal{T}_i}(\theta_i) = J^{C}_{\mathcal{T}_i}(\theta_i) - c_{1} J^{\rm VF}_{\mathcal{T}_i}(\theta_i),}
\end{equation}
where $c_1$ is the coefficient of value function loss. 

According to the target of gradient-based MRL defined in Eq. (\ref{MRL-obj}) and our objective function given by Eq. (\ref{ppo_target}), the ``outer loop'' training target of MRLCO is expressed as
\begin{equation}
\label{MRLCO_obj}
\begin{aligned}
& J^{\rm MRLCO}(\theta) = \mathbb{E}_{\mathcal{T}_i \sim \rho(\mathcal{T}), \tau \sim  P_{\mathcal{T}_i}(\tau, \theta'_i)} [J^{\rm PPO}_{\mathcal{T}_i}(\theta'_i)], \\
& {\rm where} \ \ \  \theta'_i = U_{ \tau \sim P_{\mathcal{T}_i}( \mathbf{\tau}, \theta_i)}( \theta_i, \mathcal{T}_i), \ \ \  \theta_i = \theta.
\end{aligned}
\end{equation}
Next, we can conduct gradient ascent to maximize the $J^{\rm MRLCO}(\theta)$. However, optimizing this objective function involves gradients of gradients, which introduces large computation cost and implementation difficulties when combining a complex neural network such as the seq2seq neural network. To address this challenge, we use the first-order approximation to replace the second-order derivatives as suggested in \cite{nichol2018first}, which is defined as
\begin{equation}
\label{MRLCO_grad}
g^{\rm MRLCO} := \frac{1}{n} \sum_{i=1}^{n} \left [ (\theta'_i - \theta) / \alpha / m \right ], 
\end{equation}
where $n$ is the number of sampled learning tasks in the ``outer loop'', $\alpha$ is the learning rate of the ``inner loop'' training, and $m$ is the conducted gradient steps for the ``inner loop'' training. 

We present the overall design of the algorithm in Algorithm \ref{mrlco_algo}. The parameters of the meta policy neural network are denoted as $\theta$. We firstly sample a batch of learning tasks $\mathcal{T}$ with batch size $n$ and conduct ``inner loop'' training for each sampled learning task. After finishing the  ``inner loop'' training, we update the meta-policy parameters $\theta$ by using gradient ascent $\theta \leftarrow \theta + \beta g^{\rm MRLCO}$ via \textit{Adam} \cite{kingma2014adam}. Here, $\beta$ is the learning rate of ``outer loop'' training. 

\begin{table}[t]
\renewcommand\arraystretch{1.2}
 \renewcommand{\tabcolsep}{1.5mm}
 \caption{The Neural Network and Training Hyperparameters}
 \label{tb:training_para}
 \centering
 \begin{tabular}{ c | c || c | c }
  \hline
  \textbf{Hyperparameter} & \textbf{Value} & \textbf{Hyperparameter} & \textbf{Value} \\ 
  \hline
  Encoder Layers & $2$ & Encoder Layer Type & LSTM \\
  \hline
  Encoder Hidden Units & $256$ & Encoder Layer Norm. & On \\
  \hline
  Decoder Layers & $2$ & Decoder Layer Type & LSTM \\
  \hline
  Decoder Hidden Units & $256$ & Decoder Layer Norm. & On \\
  \hline
  Learning Rate $\beta$ & $5 \times 10^{-4}$ & Learning Rate $\alpha$ & $5 \times 10^{-4}$ \\
  \hline
  Optimization Method & Adam &  Activation function & Tanh \\
  \hline
  Discount Factor $\gamma$ & $0.99$ & Loss Coefficient $c_1$ & $0.5$ \\
  \hline
  Adv. Discount Factor $\lambda$ & $0.95$ & Clipping Constant $\epsilon$ & $0.2$ \\
  \hline
  Gradient Step $m$ & $3$ &  &  \\
  \hline
 \end{tabular}
\end{table}

\section{Performance Evaluation}
\label{sec::experiments} 
This section presents the experimental results of the proposed method. First, we introduce the algorithm hyperparameters of MRLCO and the simulation environment. Next, we evaluate the performance of MRLCO by comparing it with a fine-tuning DRL method and a heuristic algorithm.  

\subsection{Algorithm Hyperparameters}
The MRLCO is implemented via Tensorflow. The encoder and decoder of the seq2seq neural network are both set as two-layer dynamic Long Short-Term Memory (LSTM) with 256 hidden units at each layer. Moreover, the layer normalization \cite{ba2016layer} is added in both the encoder and decoder. For the training hyperparameters setting in MRLCO, the learning rate of ``inner loop'' and ``outer loop'' are both set as $5 \times 10^{-4}$. The coefficient $c_1$ is set as $0.5$ and the clipping constant $\epsilon$ is set as $0.2$. The discount factors $\gamma$ and $\lambda$ are set as $0.99$ and $0.95$, respectively. The number of gradient steps for ``inner loop'' training, $m$, is set as 3. Overall, we summarize the hyperparameter setting in Table \ref{tb:training_para}.

\begin{figure}[t]
    \centering
    \includegraphics[width=3.0in]{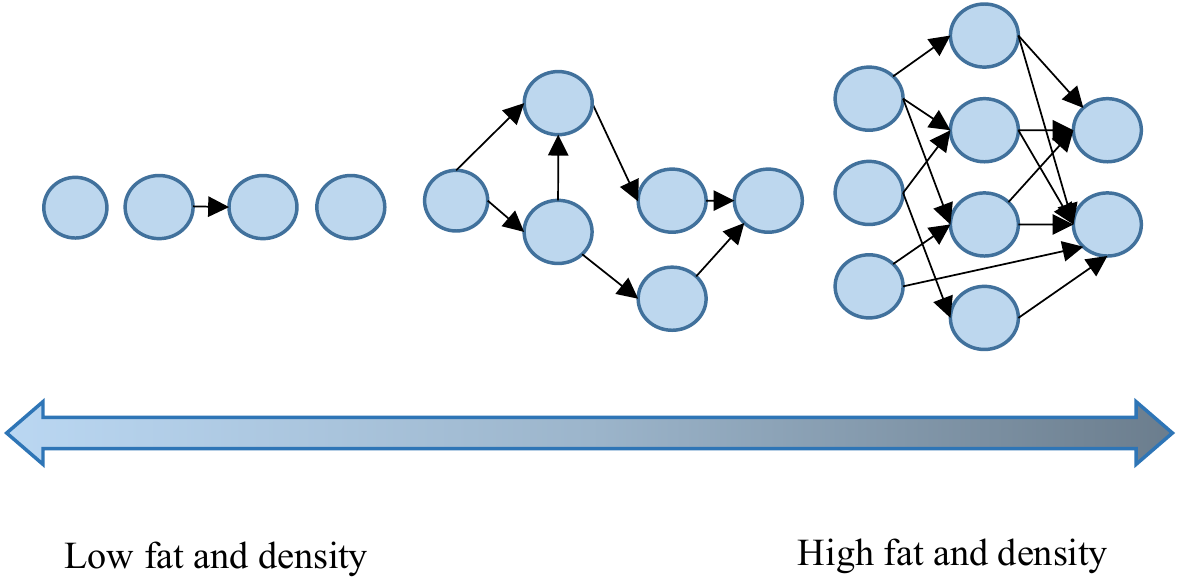}
    \centering\caption{Examples of generated DAGs. }
    \label{example_generated_dag}
\end{figure}

\begin{figure*}[!t]
\centering
\subfloat[Topology I]{\includegraphics[width=0.28\textwidth]{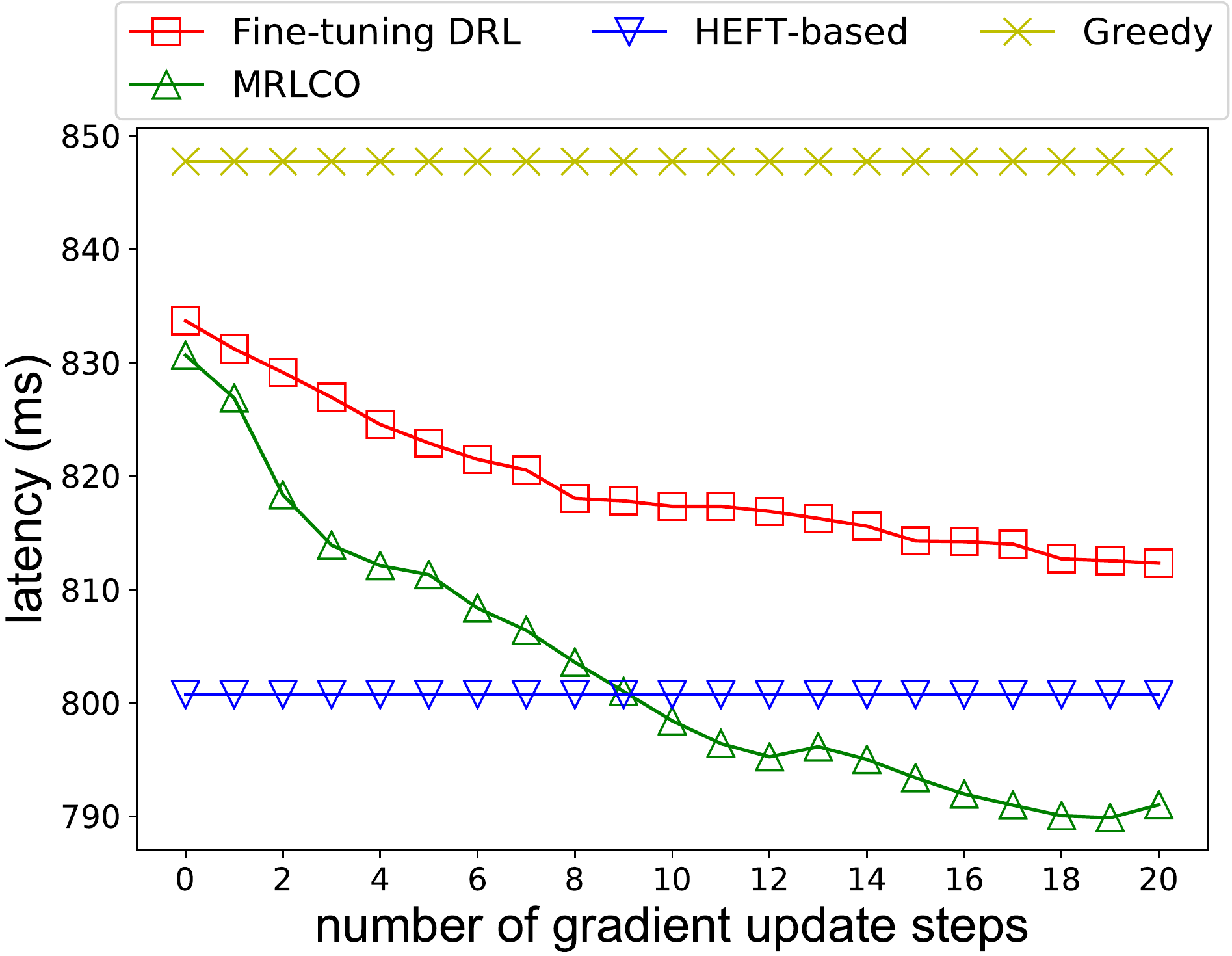}
\label{fig:toplo1}}
\hfil
\subfloat[Topology II]{\includegraphics[width=0.28\textwidth]{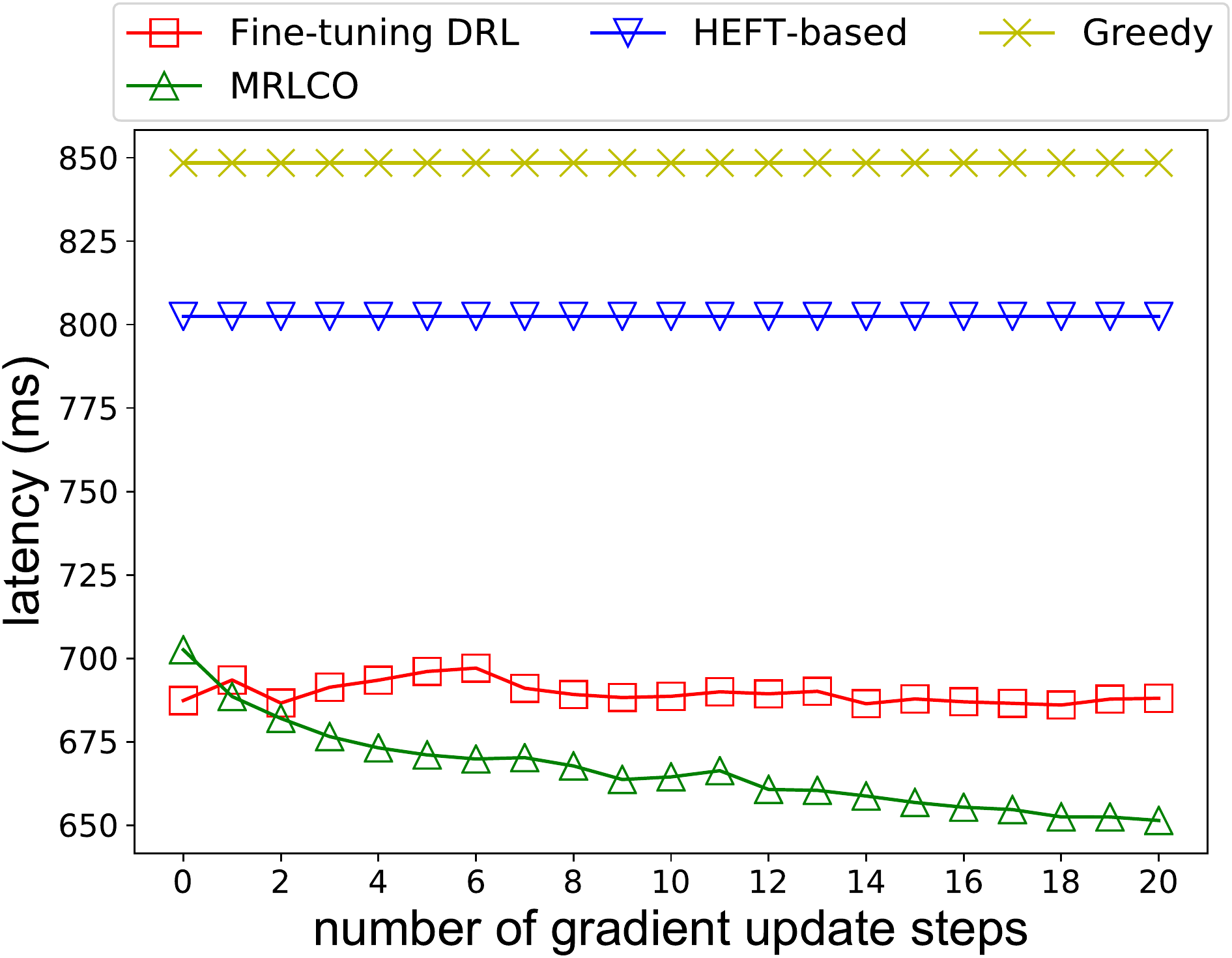}
\label{fig:toplo2}}
\hfil
\subfloat[Topology III]{\includegraphics[width=0.28\textwidth]{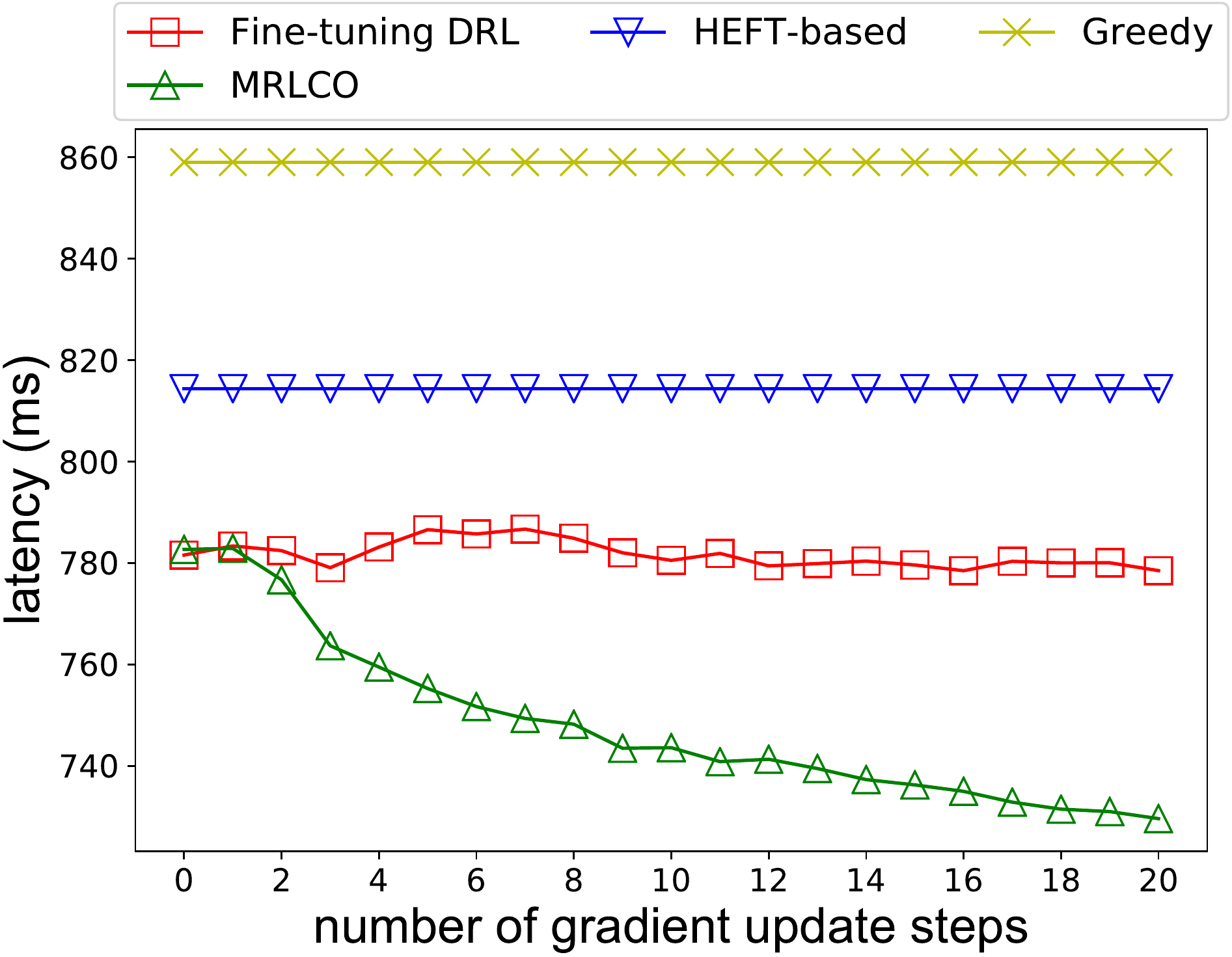}
\label{fig:toplo3}}
\caption{Evaluation results with different DAG topologies.}
\label{fig:different_toplo}
\end{figure*}

\begin{figure*}[!t]
\centering
\subfloat[$n=20$]{\includegraphics[width=0.28\textwidth]{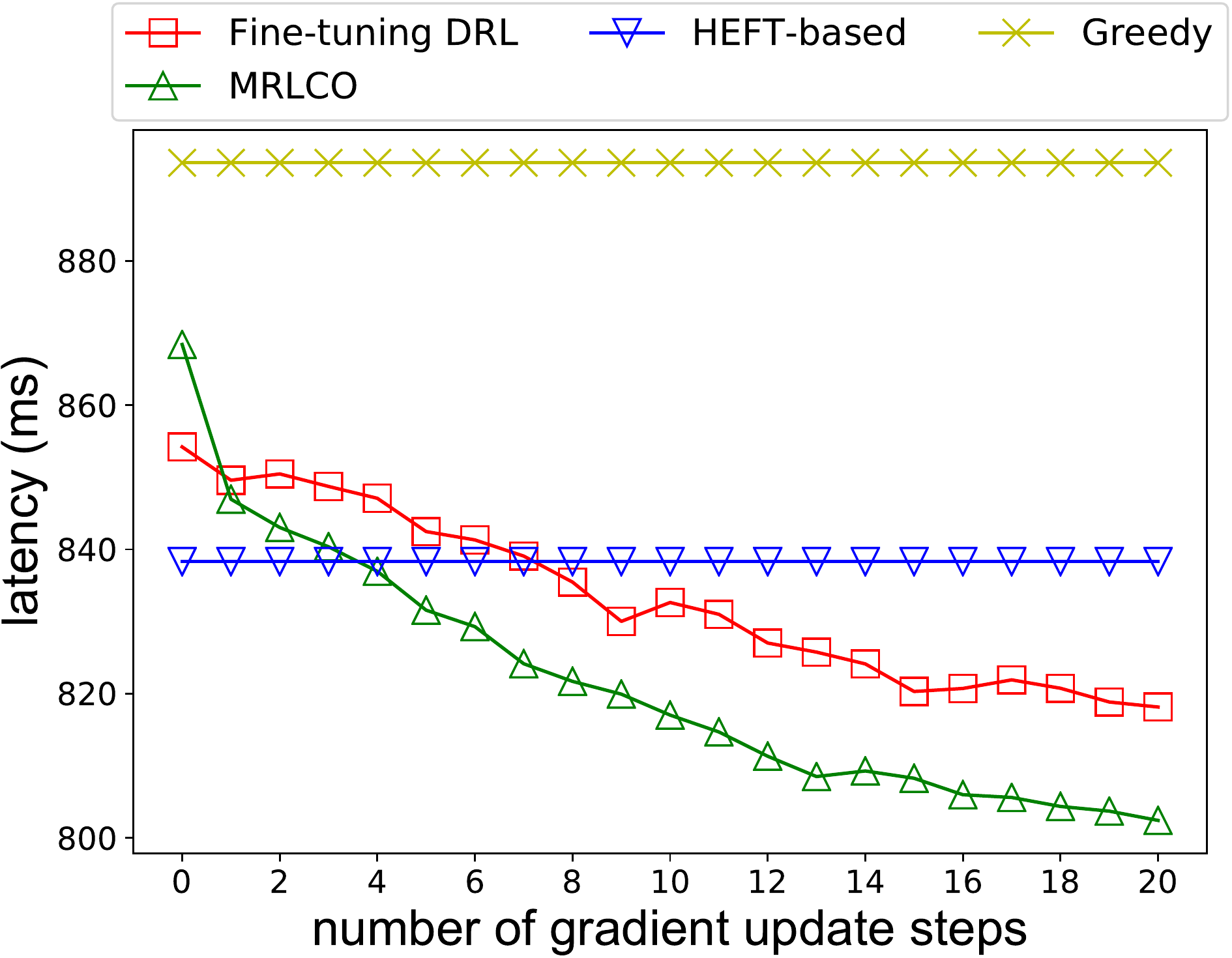}
\label{fig:n20}}
\hfil
\subfloat[$n=30$]{\includegraphics[width=0.28\textwidth]{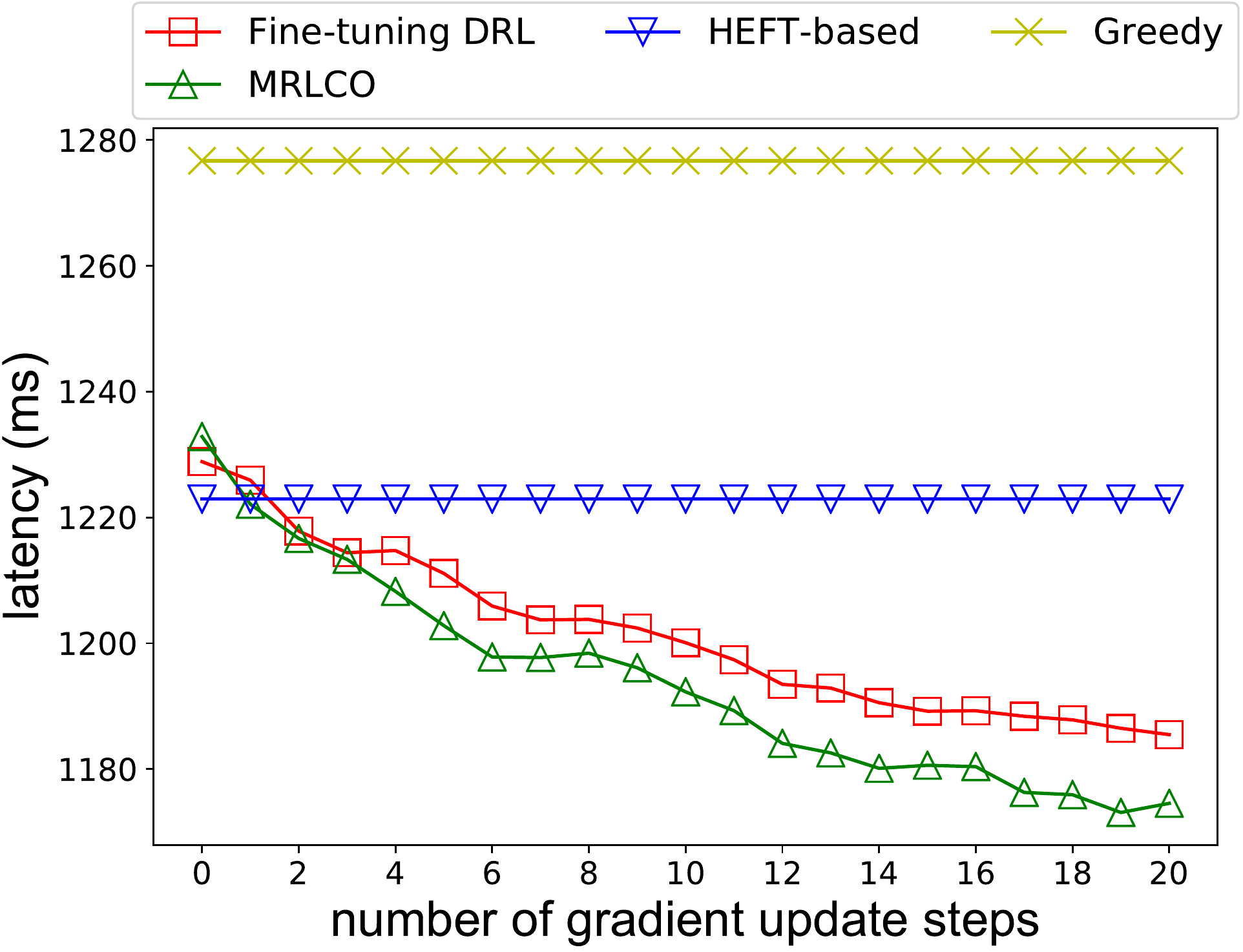}
\label{fig:n30}}
\hfil
\subfloat[$n=40$]{\includegraphics[width=0.28\textwidth]{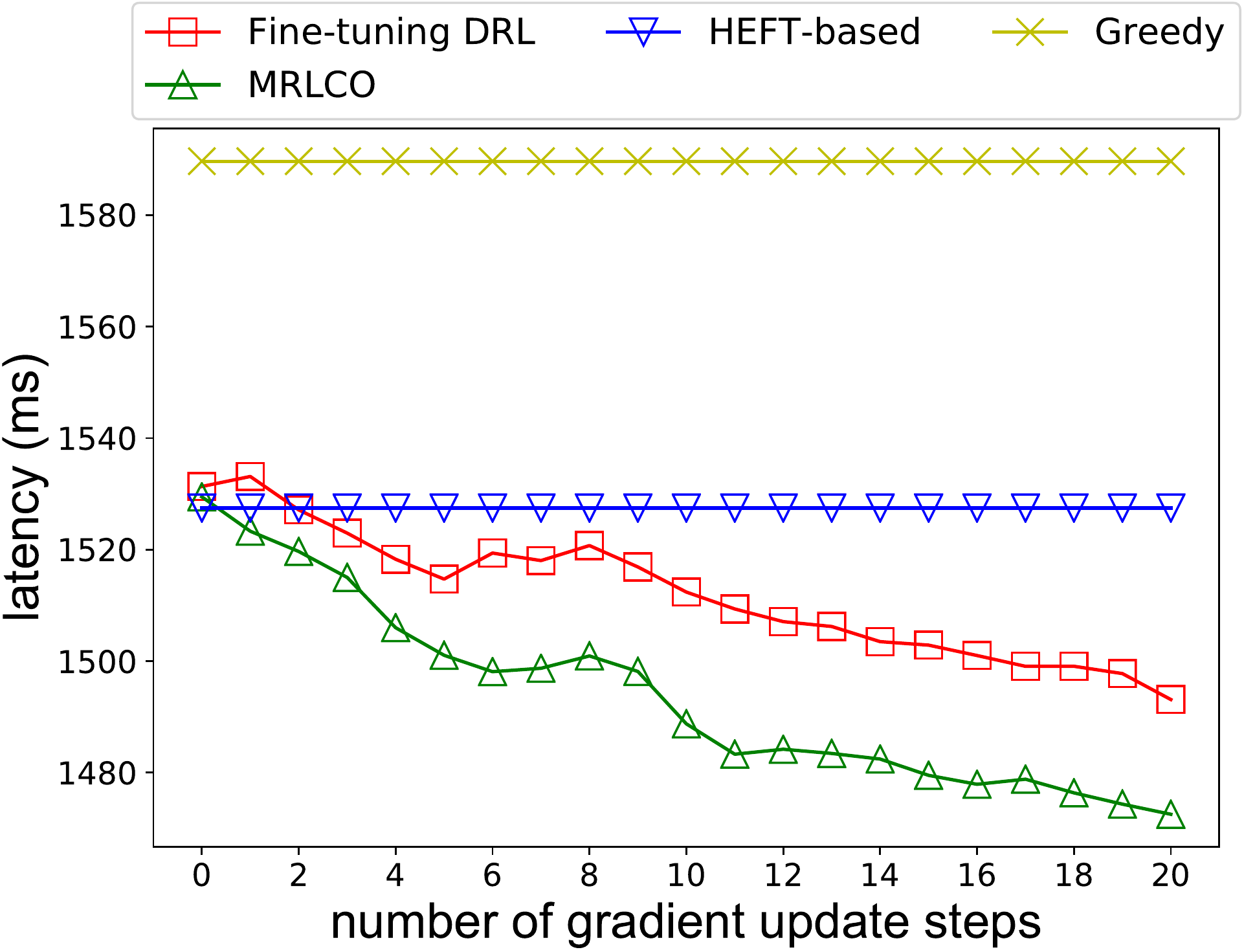}
\label{fig:n40}}
\caption{Evaluation results with different task numbers.}
\label{fig:different_n}
\end{figure*}

\subsection{Simulation Environment}
We consider a cellular network, where the data transmission rate varies with the UE's position. The CPU clock speed of UE, $f_{\rm UE}$, is set to be 1 GHz. There are 4 cores in each VM of the MEC host with the CPU clock speed of 2.5 GHz per core. The offloaded tasks can run in parallel on all cores, thus the CPU clock speed of a VM, $f_{\rm vm}$, is $4 \times 2.5 = 10$ GHz. 

Many real-world applications can be modelled by DAGs, with various topologies and task profiles. To simulate the heterogeneous DAGs, we implement a synthetic DAG generator according to \cite{arabnejad2013list}. There are four parameters controlling topologies and task profiles of the generated DAGs: \textit{n}, \textit{fat}, \textit{density}, and \textit{ccr}, where \textit{n} represents the task number, \textit{fat} controls the width and height of the DAG, \textit{density} decides the number of edges between two levels of the DAG, and \textit{ccr} denotes the ratio between the communication and computation cost of tasks. Fig. \ref{example_generated_dag} shows the generated DAGs from low fat and density to high fat and density examples. 

We design three experiments to evaluate the performance of MRLCO under dynamic scenarios. The first two experiments simulate the scenarios where UE has different application preferences represented by various topologies and task numbers. While the third experiment simulates the scenarios where UE has varying dynamic transmission rates. For all experiments, the data size of each task ranges from 5 KB to 50 KB; the CPU cycles required by each task ranges from $10^7$ to $10^8$ cycles \cite{dinh2017offloading}. The length of child/parent task indices vector $p$ is set as 12. We randomly select \textit{ccr} from 0.3 to 0.5 for each generated DAG, since most of mobile applications are computation-intensive. The generated datasets in each experiment are separated into ``training datasets'' and ``testing datasets''. We consider learning an effective offloading policy for each dataset as a learning task. The MRLCO firstly learns a meta policy based on ``training datasets'' by using Algorithm \ref{mrlco_algo}. The learned meta policy is then used as the initial policy to fast learn an effective offloading policy for the ``testing datasets''. 

We compare MRLCO with three baseline algorithms: 
\begin{itemize}{}
\item \textit{Fine-tuning DRL}: It first pretrains one policy for all ``training datasets'' using the DRL-based offloading algorithm proposed in \cite{wang2019computation}. Next, it uses the parameters of the trained policy network as an initial value of the task-specific policy network, which is then updated on the ``testing datasets''.  
\item \textit{HEFT-based}: This algorithm is adapted from \cite{lin2015task}, which firstly prioritizes tasks based on HEFT and then schedules each task with earliest estimated finish time. 
\item \textit{Greedy}: Each task is greedily assigned to the UE or the MEC host based on its estimated finish time. 
\end{itemize}

\subsection{Results Analysis}

\begin{figure*}[!t]
\centering
\subfloat[$R_{\rm ul} = R_{\rm dl} = 5.5$ Mbps]{\includegraphics[width=0.28\textwidth]{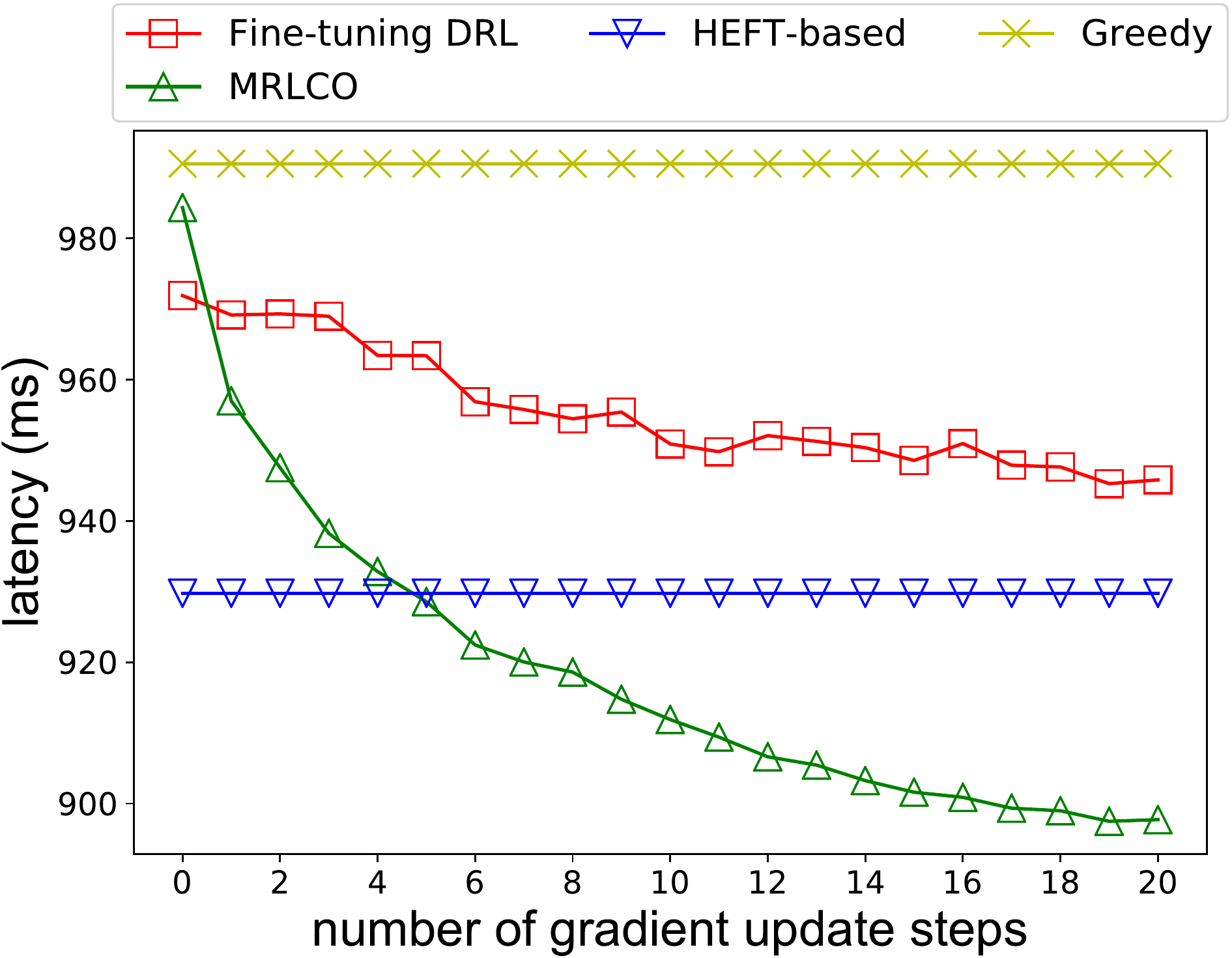}
\label{fig:transrate5.5}}
\hfil
\subfloat[$R_{\rm ul} = R_{\rm dl} = 8.5$ Mbps]{\includegraphics[width=0.28\textwidth]{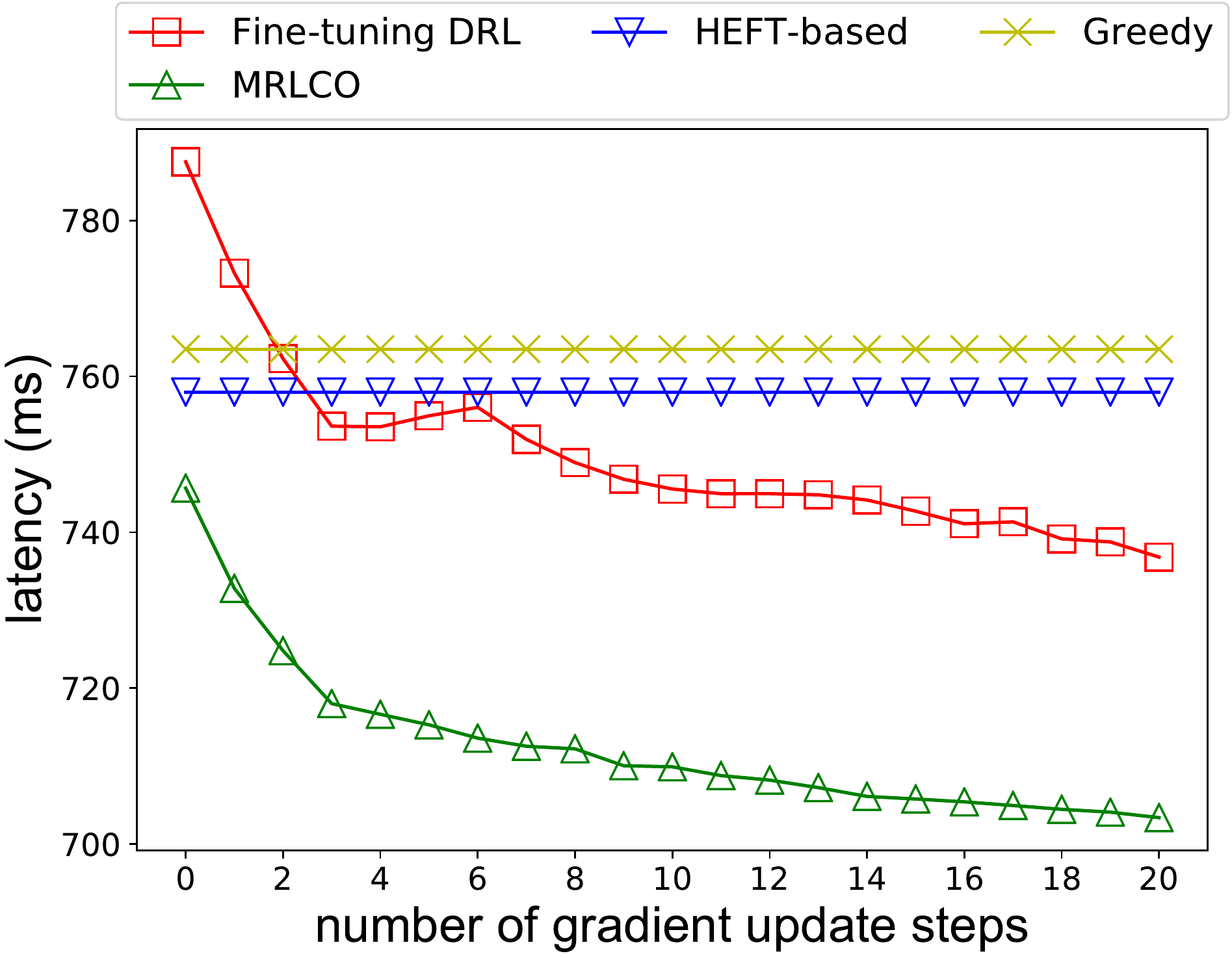}
\label{fig:transrate8.5}}
\hfil
\subfloat[$R_{\rm ul} = R_{\rm dl} = 11.5$ Mbps]{\includegraphics[width=0.28\textwidth]{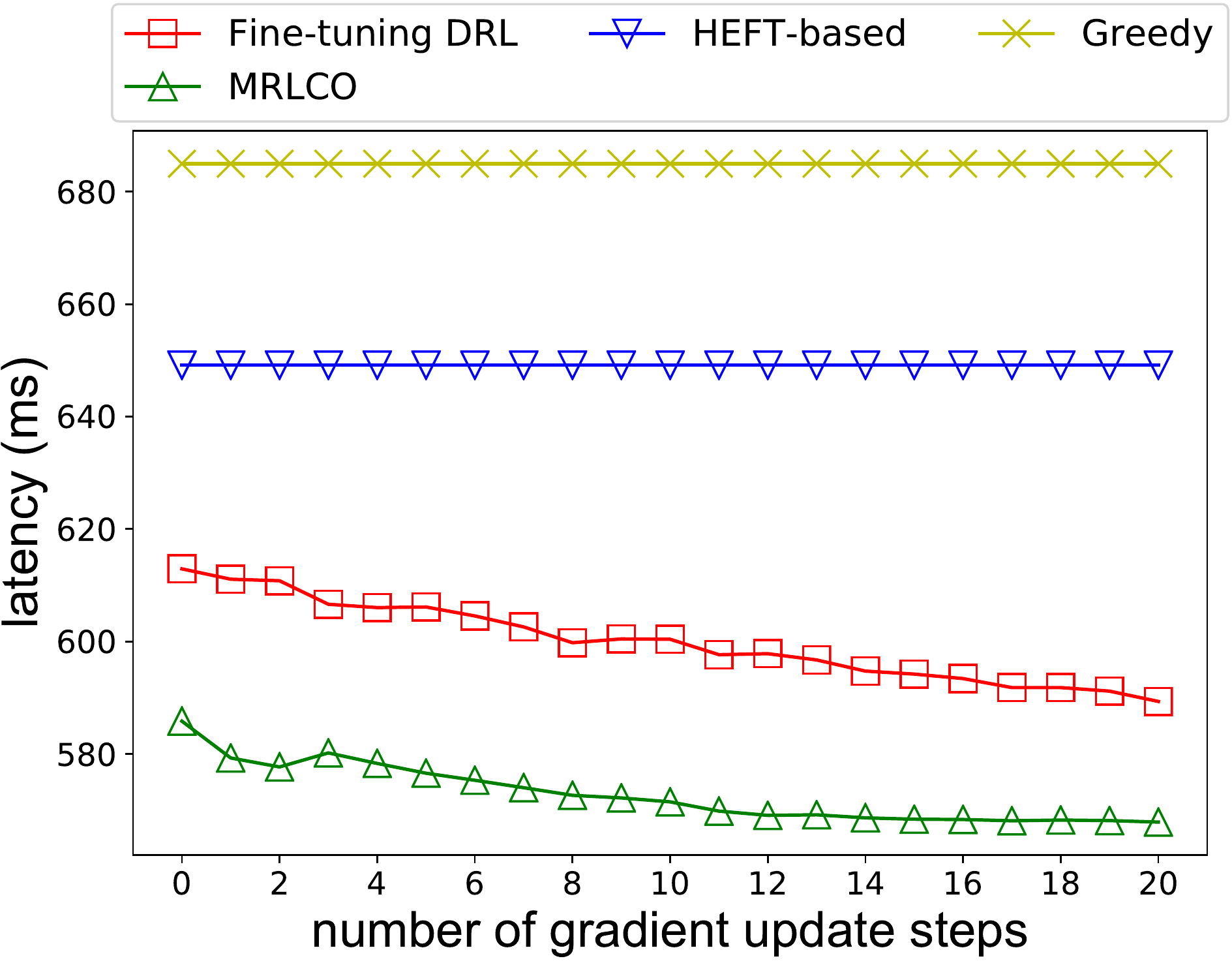}
\label{fig:transrate11.5}}
\caption{Evaluation results with different transmission rates.}
\label{fig:different_trans_rate}
\end{figure*}

\begin{table*}[ht]
\renewcommand\arraystretch{1.2}
  \begin{center}
    \caption{The comparison of MRLCO and baseline algorithms in average latency ({\rm ms}) on different testing datasets,
    \protect\\ N/A denotes cases unable to find optimum.}
    \label{avg_latency_table}
    \begin{tabular*}{\textwidth}{c*{6}{c}cccc} 
      \hline
      \ \makebox[6em]{\textbf{Testing dataset}} \ & & \multicolumn{2}{c}{\textbf{Heuristic Algorithms}} & &\multicolumn{2}{c}{\textbf{Fine-tuning DRL}} & & &\multicolumn{2}{c}{\textbf{MRLCO}} \\ 
      \cline{3-4} \cline{6-7} \cline{9-10}
      & \makebox[4em]{Optimal} & \makebox[4em]{HEFT-based} & \makebox[4em]{Greedy} & & \makebox[6.5em]{update steps(20)} & \makebox[6.5em]{update steps(100)} & &\makebox[6.5em]{update steps(20)} & \makebox[6.5em]{update steps(100)}\\
      \hline
      \ Toploogy I \  & 679.31 & 800.75 & 847.73 & & 812.32 & 789.92 & & 791.03 & 722.63\\
      \ Toploogy II \  & 555.46 & 802.46 & 848.43 & & 688.05 & 636.49 & & 651.42 & 601.93\\
      \ Toploogy III\  & 605.05 & 814.39 & 859.03 & & 778.52 & 712.79 & & 729.63 & 641.92\\
      \ $n = 20$ \  & 689.21 & 838.31 & 893.62 & & 818.14 & 802.50 & & 802.41 & 743.42\\
      \ $n = 30$ \  & N/A & 1222.93 & 1276.70 & & 1185.47 & 1152.07 & & 1174.55 & 1098.43\\
      \ $n = 40$ \  & N/A & 1527.47 & 1589.66 & & 1493.11 & 1432.41 & & 1472.53 & 1397.63\\
      \ $R_{\rm ul} = R_{\rm dl} = 5.5$ Mbps  & 770.10 & 929.79 & 990.58 & & 945.82 & 901.36 & & 897.73 & 831.58\\
      \ $R_{\rm ul} = R_{\rm dl} = 8.5$ Mbps & 628.21 & 757.99 & 763.49 & & 736.81 & 701.75 & & 703.38 & 674.93\\
      \ $R_{\rm ul} = R_{\rm dl} = 11.5$ Mbps & 524.14 & 649.15 & 684.97 & & 589.33 & 570.19 & & 567.88 & 548.26\\
      \hline
    \end{tabular*}
  \end{center}
\end{table*}

In the first experiment, we generate DAG sets with different topologies to simulate the scenario where users have different preferences of mobile applications. Each dataset contains 100 DAGs of similar topologies with the same \textit{fat} and \textit{density}, which are two key parameters influencing the DAG topology. We set the task number for each generated DAG as $n=20$ and set $\textit{fat} \in \{0.4, 0.5, 0.6, 0.7, 0.8\}$, $\textit{density} \in \{0.4, 0.5, 0.6, 0.7, 0.8\}$. 25 DAG sets are generated with different combinations of fat and density. Each DAG set represents the application preference of one mobile user and consider finding the effective offloading policy for a DAG set as a learning task. We randomly select 22 DAG sets as the training datasets and the other 3 as unseen testing datasets. We train the MRLCO and the fine-tuning DRL method on the training datasets and evaluate MRLCO and baseline algorithms on the testing datasets. 

During training of MRLCO, we set the meta batch size as 10, thus 10 learning tasks are sampled from $\rho(\mathcal{T})$ in the ``outer loop'' training stage. At each ``inner loop'', we sample 20 trajectories for a DAG and conduct $m$ policy gradient updates ($m=3$) for the PPO target. After training, we evaluate the MRLCO and fine-tuning DRL method by running up to 20 policy gradient updates, each samples 20 trajectories for a DAG on the testing datasets. Fig. \ref{fig:different_toplo} shows the performance of the MRLCO and baseline algorithms with different DAG sets. Overall, the Greedy algorithm has the highest latency, while the MRLCO obtains the lowest latency. Fig. \ref{fig:toplo1} demonstrates that the MRLCO is better than the HEFT-based algorithm after 9 steps of gradient update, while the fine-tuning DRL method consistently performs worse than the HEFT-based algorithm. This indicates that the MRLCO can adapt to new tasks much more quickly than the fine-tuning DRL method. In Fig. \ref{fig:toplo2} and Fig. \ref{fig:toplo3}, the MRLCO and the fine-tuning DRL method with 0 step of gradient updates already beat the two heuristic-based algorithms: HEFT-based and the Greedy algorithms, {\color{black} because both the MRLCO and fine-turning DRL learn the updated policy based on the pre-trained models instead of learning from scratch.} These heuristic-based algorithms use fixed policies to obtain the offloading plan, which cannot adapt well to different DAG topologies. 

The second experiment aims to show the influence of the task number $n$ on the performance of different algorithms. We randomly generate 6 training datasets with $n \in \{10, 15, 25, 35, 45, 50\}$ and 3 testing datasets with $n \in \{20, 30, 40\}$. In each dataset, we generate DAGs by randomly selecting \textit{fat} from $\{0.4, 0.5, 0.6, 0.7, 0.8\}$, \textit{density} from $\{0.4, 0.5, 0.6, 0.7, 0.8\}$, and \textit{ccr} from 0.3 to 0.5, thus the distributions of DAG topologies of all datasets are similar. In this experiment, we set the meta batch size as 5 and the rest of the settings the same as the first experiment. Fig. \ref{fig:different_n} shows that both the MRLCO and the fine-tuning DRL method outperform the HEFT-based algorithms after a few gradient updates, and are consistently better than the Greedy from step 0 of gradient updates. Moreover, MRLCO adapts to new learning tasks faster than the fine-tuning DRL method. For example, Fig. \ref{fig:n30} shows that, after one step gradient update, the latency of MRLCO decreases sharply and is less than both fine-tuning and HEFT-based algorithms. After 20 gradient updates, MRLCO obtains the lowest latency compared to the baseline algorithms. 

We conduct the third experiment to evaluate the performance of MRLCO with different transmission rates. Learning the offloading policy for each transmission rate is considered as an individual learning task. We randomly generate the DAG dataset by setting $n=20$ and other parameters the same as the second experiment. In addition, we implement \textit{Optimal} algorithm via exhaustively searching the solution space to find the optimal offloading plan. We conduct meta training process based on randomly selected transmission rates from 4 Mbps to 22 Mbps with a step size of 3 Mbps. We then evaluate the trained meta policy among transmission rates from \{5.5 Mbps, 8.5 Mbps, 11.5 Mbps\}, which are unseen in the training procedure. Fig. \ref{fig:different_trans_rate} shows that the MRLCO again adapts to new learning tasks much faster than the fine-tuning DRL method in all test sets and achieves the lowest latency after 20 gradient updates. In some cases (Fig. \ref{fig:transrate8.5} and Fig. \ref{fig:transrate11.5}), MRLCO even achieves the lowest latency at the initial point. 

Table \ref{avg_latency_table} summarizes the average latency of all algorithms on different testing datasets. Overall, the MRLCO outperforms all heuristic baseline algorithms after 20 gradient update steps. The MRLTO and fine-tuning DRL method will get better results with more update steps. Table \ref{avg_latency_table} also shows the performance of the fine-tuning and the MRLCO algorithms after 100 update steps. Compared to the fine-tuning algorithm, the MRLCO achieves better result after both 20 and 100 update steps. However, there are still gaps between the results of MRLCO and the Optimal values. One possible solution could be to integrate the seq2seq neural network with another sample efficient off-policy MRL method \cite{rakelly2019efficient}, which is a direction for future work.

\section{Related Work}
\label{sec::relatedwork}





The task offloading problem in MEC has attracted significant research interests \cite{dinh2017offloading,huang2019deeprl,dinh2018learning,chen2019optimized,mao2017survey,chen2018task,hong2018qoe,wang2017online,neto2018uloof,zanni2017automated,zanni2017automated_info,zhan2020deep,tan2018mobility,huang2019deep,ning2019deep,zhan2020mobility}. In general, there are two task models used in the related work: task model for binary offloading and that for partial offloading \cite{mao2017survey}. In the task model for binary offloading, there are no inner dependencies among computation tasks for an application. Dinh et al. \cite{dinh2017offloading} aimed to find an offloading plan for a set of tasks among different access points and MEC hosts, in order to achieve the minimal joint target of latency and energy. Chen et al. \cite{chen2018task} focused on computation offloading for independent tasks in a software-defined ultra-dense network. They formulated the task offloading problem as a mixed integer non-linear program and solved it by using decomposition and heuristic methods. Hong et al. \cite{hong2018qoe} proposed an approximate dynamic programming algorithm for computation offloading to achieve the optimal quality of experience. In the task model for partial offloading, applications were composed of tasks with inner dependencies, which is able to achieve a fine granularity of computation offloading, leading to better offloading performance. Wang et al. \cite{wang2017online} modelled both the applications and the computing system as graphs and proposed an approximation algorithm for finding the task offloading plan to obtain the lowest cost. {\color{black} Neto et al. \cite{neto2018uloof} implemented a user-level online offloading framework for Android applications, aiming at minimizing the remote execution overhead. Zanni et al. \cite{zanni2017automated} proposed an innovative task selection algorithm for Android applications, achieving method-level granularity of offloading. }

In order to adapt the offloading strategies for dynamic scenarios, over recent years, DRL has been widely applied to solve task offloading problems in MEC systems. Dinh et al. \cite{dinh2018learning} focused on the multi-user multi-edge-node computation offloading problem by using deep Q-learning. Chen et al. \cite{chen2019optimized} considered an ultra-dense network, where multiple base stations can be selected for offloading. They also adopted deep Q-learning to obtain the offloading strategy. Huang et al. \cite{huang2019deep} proposed a DRL-based offloading framework which jointly considers both the offloading decisions and resource allocations. {\color{black} Zhan et al. \cite{zhan2020deep} proposed an efficient task offloading method combining PPO and convolutional neural networks. } Tan et al. \cite{tan2018mobility} proposed a deep Q-learning based offloading method considering constraints of limited resources, vehicle's mobility and delay. Huang et al. \cite{huang2019deeprl} proposed a DRL-based online offloading framework to maximize the weighted sum of the computation rates of all the UE. {\color{black} Ning et al. \cite{ning2019deep} proposed a deep Q-learning based method for jointly optimising task offloading and resource allocation in MEC.} The existing studies mostly assume the offloading problem as one learning task and apply conventional DRL algorithms to solve the task. However, many DRL algorithms suffer from poor sample efficiency --- when facing new scenarios, those DRL-based offloading methods need long training time to find the effective policy, which impedes their practical deployment. To address this issue, our offloading method adopts an MRL approach which can efficiently solve new learning tasks with the requirement of only few gradient update steps and a small amount of data. As a result, our method can quickly adapt to the changes of environments with the requirement of only few training steps rather than fully retraining the offloading policy from scratch. With the low demands of computation and data, our method can efficiently run on the resource-constrained UE using its own data. 

\section{Discussion}
\label{sec::disscussion}

{\color{black} MRLCO has many advantages over the existing RL-based task offloading methods, such as learning to fast adapt in a dynamic environment and high sample efficiency. Beyond the scope of task offloading in MEC systems, the proposed MRLCO framework has the potential to be applied to solve more decision-making problems in MEC systems. For instance, content caching in MEC aims to cache popular contents at MEC hosts to achieve high Quality-of-Service (QoS) for mobile users and reduce the network traffic. While MEC hosts can have different caching policies to suit the dynamic content preferences and network conditions of users in different areas. The proposed MRL framework can be adapted to solve this problem through executing the ``outer loop'' training at cloud servers to learn a meta caching policy and the ``inner loop'' training at MEC hosts to learn a specific caching policy for each MEC host. 

Even though MRLCO has many benefits to MEC systems, there are several challenges for further exploration. In this paper, we consider stable wireless channels, reliable mobile devices, and sufficient computation resources. Thus, the MRLCO will not break down when increasing the number of users.  However, when operating at large-scale, some UE as stragglers may drop out due to broken network connections or insufficient power. Considering the synchronous process of ``outer loop'' training that updates the meta policy after gathering parameters from all UE, the stragglers might affect the training performance of MRLCO. One way to solve this issue is to apply an adaptive client selection algorithm which can automatically filter out stragglers and select reliable clients to join the training process based on their running states. }

\section{Conclusion}
\label{sec::conclusion}
This paper proposes an MRL-based approach, namely MRLCO, to solve the computation offloading problem in MEC. Distinguished from the existing works, the MRLCO can quickly adapt to new MEC environments within a small number of gradient updates and samples. In our proposed method, the target mobile applications are modelled as DAGs, the computation offloading process is converted to a sequence prediction process, and a seq2seq neural network is proposed to effectively represent the policy. Moreover, we adopt the first-order approximation for the MRL objective to reduce the training cost and add a surrogate clipping to the objective so as to stabilize the training. We conduct simulation experiments with different DAG topologies, task numbers, and transmission rates. The results demonstrate that, within a small number of training steps, MRLCO achieves the lowest latency  compared to three baseline algorithms including a fine-tuning DRL method, a greedy algorithm, and an HEFT-based algorithm.

\bibliographystyle{IEEEtran} 
\bibliography{metaRL-task-offloading}

\end{document}